\documentclass[12pt]{article}
\pdfoutput=1
\usepackage{epsfig}
\usepackage{amsfonts}
\usepackage{amscd}
\usepackage{latexsym}
\usepackage{amsmath,amssymb}
\usepackage{verbatim}
\usepackage{setspace}
\usepackage{color}
\usepackage{fancyhdr}
\usepackage{cite}
\usepackage{hyperref}
\usepackage{tikz}
\usepackage{slashed}
\usetikzlibrary{calc}   
\usetikzlibrary{topaths}
\usetikzlibrary{decorations}
\usetikzlibrary{decorations.pathmorphing}

\usepackage{draft} 
\usepackage{hyperref}
\usepackage{graphicx,color,subfig}
\usepackage{cite}
\usepackage{mciteplus}
\usepackage{skak}
\DeclareFontFamily{OT1}{pzc}{}
\DeclareFontShape{OT1}{pzc}{m}{it}{<-> s * [1.10] pzcmi7t}{}
\DeclareMathAlphabet{\mathpzc}{OT1}{pzc}{m}{it}

\numberwithin{equation}{section}

 \def\p{\partial}

\def\0{{(0)}}
\def\1{{(1)}}
\def\2{{(2)}}

\def\<{\langle }
\def\>{\rangle }

\def\eads${H$_3$}

\newcommand{\ba}{\begin{align}}
\newcommand{\ea}{\end{align}}
\def\be{\begin{equation}}
\def\ee{\end{equation}}
\def\beq{\be\begin{array}{c}}
\def\eeq{\end{array}\ee}

\def\be#1\ee{\begin{align}#1\end{align}}

\begin{document}

\newcommand{\scriplus}{\mathcal{I}^+}
\newcommand{\Del}{\nabla}

	 \renewcommand{\theequation}{\thesection.\arabic{equation}}
   \makeatletter
  \let\over=\@@over \let\overwithdelims=\@@overwithdelims
  \let\atop=\@@atop \let\atopwithdelims=\@@atopwithdelims
  \let\above=\@@above \let\abovewithdelims=\@@abovewithdelims
\renewcommand\section{\@startsection {section}{1}{\z@}%
                                   {-3.5ex \@plus -1ex \@minus -.2ex}
                                   {2.3ex \@plus.2ex}%
                                   {\normalfont\large\bfseries}}

\renewcommand\subsection{\@startsection{subsection}{2}{\z@}%
                                     {-3.25ex\@plus -1ex \@minus -.2ex}%
                                     {1.5ex \@plus .2ex}%
                                     {\normalfont\bfseries}}

\renewcommand{\H}{\mathcal{H}}
\newcommand{\SU}{\mbox{SU}}
\newcommand{\chiu}{\chi^{{\rm U}(\infty)}}
\newcommand{\ff}{\rm f}
\linespread{1.3}

\unitlength = .8mm

\begin{titlepage}

\begin{center}

\hfill \\
\hfill \\
\vskip 1cm

\title{A Conformal Basis for 
Flat Space Amplitudes}

\author{Sabrina Pasterski$^1$ and Shu-Heng Shao$^2$}

\address{
$^1$Center for the Fundamental Laws of Nature, Harvard University,\\
Cambridge, MA 02138, USA
\\
$^2$School of Natural Sciences, Institute for Advanced Study, \\Princeton, NJ 08540, USA
}

\end{center}

\vspace{2.0cm}

\begin{abstract}
We study solutions of the  Klein-Gordon,  Maxwell, and linearized Einstein equations in $\mathbb{R}^{1,d+1}$ that transform as $d$-dimensional conformal primaries under the Lorentz group $SO(1,d+1)$.  Such solutions, called conformal primary wavefunctions, are labeled by a conformal dimension $\Delta$ and a point in $\mathbb{R}^d$, rather than an  on-shell $(d+2)$-dimensional momentum.   
 We show that  the continuum of  scalar conformal primary wavefunctions on the principal continuous series  $\Delta\in \frac d2+ i\mathbb{R}$ of $SO(1,d+1)$ spans a complete set  of normalizable solutions to the wave equation.   In the massless case, with or without spin, the transition from momentum space to conformal primary wavefunctions is implemented by a  Mellin  transform.  
 As a consequence of this construction, scattering amplitudes in this basis transform covariantly under $SO(1,d+1)$  as $d$-dimensional conformal correlators.  

\end{abstract}

\vfill

\end{titlepage}

\eject
\tableofcontents

\section{Introduction}

Scattering problems are conventionally studied in  momentum space where translation symmetry is manifest.  
However, not all properties of  scattering amplitudes are emphasized in this choice of basis. One famous alternative basis is  twistor space \cite{penrose1967twistor,Witten:2003nn} where many remarkable properties of perturbative gauge theory amplitudes are naturally explained.

The Lorentz group in $\mathbb{R}^{1,d+1}$ is identical to the Euclidean $d$-dimensional conformal group $SO(1,d+1)$.  It is then natural to ask if there is a basis of wavefunctions where scattering amplitudes in $\mathbb{R}^{1,d+1}$ admit interpretations as Euclidean $d$-dimensional conformal correlators.\footnote{By a conformal correlator we mean a function of $n$ points on $\mathbb{R}^d$ that transforms covariantly as in \eqref{conformalcorrelator} (or the spin version thereof), i.e. it is a function with the same conformal covariance as  an $n$-point function of primaries in conformal field theory (CFT).}   For massless scalar and spin-one fields, such wavefunctions were  constructed in \cite{deBoer:2003vf,Cheung:2016iub}, while the massive scalar wavefunctions were introduced in \cite{Pasterski:2016qvg}  for $(3+1)$ spacetime dimensions.  However, it had not been established whether these conformal wavefunctions form a complete set of normalizable solutions to the wave equation in each case. In this paper we study the completeness of these wavefunctions with and without spin and extend the construction to arbitrary spacetime dimensions.

The search for conformal bases of wavefunctions has its roots in the study of two-dimensional  conformal symmetries in  four-dimensional scattering amplitudes.  
In \cite{deBoer:2003vf}, de Boer and Solodukhin approached the problem  of flat space holography from a hyperbolic slicing of  Minkowski space. Since each slice is a copy of a three-dimensional hyperbolic space $H_3$, the two-dimensional conformal symmetry naturally arises  on the boundary  via $AdS$ holography.  
It was then conjectured  \cite{Banks:2003vp,Barnich:2009se,Barnich:2010eb,Barnich:2011ct} that in any four-dimensional quantum gravity, the Lorentz group $SL(2,\mathbb{C})$ is  enhanced to the full Virasoro symmetry.   
This conjecture was later refined and verified \cite{Kapec:2014opa,Kapec:2016jld,Cheung:2016iub} for tree-level $\mathcal{S}$-matrices  following a new subleading soft graviton theorem \cite{Cachazo:2014fwa}.   
In particular,   the authors of \cite{Kapec:2016jld} gave an explicit construction of a two-dimensional stress-tensor that generates a complex Virasoro symmetry acting on the celestial sphere at null infinity. 
 The one-loop correction to the Virasoro stress tensor was recently discussed in \cite{Hawking:2016sgy,He:2017fsb} from an anomaly \cite{He:2014bga,Bianchi:2014gla,Bern:2014oka,Bern:2014vva} of this subleading soft graviton theorem.  
Furthermore,  it was observed that  insertions of  soft photons in the amplitude resemble the  Kac-Moody algebra in two dimensions \cite{Strominger:2013lka,He:2015zea,Lipstein:2015rxa,Cardona:2015woa,Nande:2017dba}.      See \cite{Strominger:2017zoo} for a comprehensive review of this subject.

In this paper we introduce a basis of flat space wavefunctions  that is natural  for the study of the $d$-dimensional conformal structure of $(d+2)$-dimensional scattering amplitudes beyond the soft limit.  
We consider on-shell wavefunctions in $\mathbb{R}^{1,d+1}$ with spin that are $SO(1,d+1)$ conformal primaries, extending the construction   in \cite{Cheung:2016iub,Pasterski:2016qvg}  beyond $(3+1)$ dimensions.  These solutions, called conformal primary wavefunctions, are labeled by a conformal dimension $\Delta$ and a point $\vec w$  in $\mathbb{R}^d$, as opposed to a $(d+2)$-dimensional on-shell momentum.  Crucially, the conformal dimension $\Delta$ should be thought of as a continuous label for solutions in this basis, and is \textit{not} fixed by the mass of the bulk field.  Rather, we require a continuum of conformal primary wavefunctions to span the solution space of a single bulk field.\footnote{Therefore in any flat space holographic duality formulated through this construction, the dual putative conformal theory will be non-compact (like a Liouville theory). This point was also advocated in \cite{deBoer:2003vf,Cheung:2016iub}.}

One immediate puzzle about the massive conformal primary wavefunction is that, as opposed to the massless case, there does not appear to be a canonical way to associate a point $\vec w$ in $\mathbb{R}^d$  to the trajectory of a massive particle in $\mathbb{R}^{1,d+1}$.   It turns out that the label $\vec w$ is not a point in  position space, but  in  momentum space of the massive particle.  More precisely, $\vec w$ is a boundary point of the space of $(d+2)$-dimensional on-shell momenta, which is a copy of a $(d+1)$-dimensional hyperbolic space $H_{d+1}$.  See \eqref{CPW} for details.

The main objective of this paper is to determine the range of the conformal dimension $\Delta$ for these  conformal primary wavefunctions to form a basis of on-shell wavefunctions in $\mathbb{R}^{1,d+1}$.  If such range of $\Delta$ exists,   then we can safely translate every scattering amplitude into this conformal primary basis without loss of information.  In both the massive and the massless scalar cases, we show that the continuum of  conformal primary wavefunctions with\footnote{For massive scalar conformal primary wavefunctions, we only require half of the principal continuous series, i.e. $\Delta\in \frac d2+ i \mathbb{R}_{\ge0}$ or $\Delta\in \frac d2+ i \mathbb{R}_{\le0}$. See Section \ref{sec:basis} for more details.}
\begin{align}\label{introprincipal}
\Delta\in \frac d2+ i \mathbb{R}\,,
\end{align}
 spans the complete set of delta-function-normalizable  solutions with respect to the Klein-Gordon norm \eqref{KGip}.  This range of $\Delta$ is known as the principal continuous series of irreducible unitary representations of $SO(1,d+1)$, which plays a central role in the harmonic analysis of the conformal group (see, for example, \cite{sugiura1990unitary}).  We  contrast the scalar conformal primary basis with the momentum basis  in Tables \ref{table:massive} and \ref{table:massless}.  In the massless case, the change of basis is given by a Mellin transform (or plus a shadow transform).  In the massive, on the other hand, it is implemented by an integral over all the on-shell momenta with the  bulk-to-boundary propagator in $H_{d+1}$ being the Fourier coefficient.

\begin{table}[h!]
\centering
\begin{tabular}{|c|c|c|c|}
\hline
Bases & ~~~~Plane Waves~ ~~~
&~ Conformal Primary Wavefunctions~ \\
\hline
&&\\
~~~\text{Notations}~~~&$~  \exp\left[ \pm i p\cdot X\right]~$  
&$\phi_{\Delta }^\pm (X^\mu; \vec w)$ \\
&&\\
\hline
&&\\
Labels & $ p^\mu  ~~ (p^2=-m^2,  p^0> 0)$&$ \Delta\in\frac d2 +i \mathbb{R}_{\ge0} ~,~\vec w\in\mathbb{R}^d$\\
&&\\
\hline
\end{tabular}
\caption{A comparison between the plane wave basis and the  conformal primary basis $\mathcal{B}^\pm$ for  normalizable, outgoing/incoming solutions to the massive Klein-Gordon equation. The plane wave is labeled by  an on-shell momentum $p^\mu$ with $p^2=-m^2$, whereas the conformal primary wavefunction is labeled by $\Delta \in  \frac d2 + i\mathbb{R}_{\ge0}$ and $\vec w\in\mathbb{R}^d$.  Here the plus and minus superscripts denote outgoing and incoming wavefunctions, respectively. There is another basis $\widetilde {\mathcal{B}}^\pm$  that is shadow to $\mathcal{B}^\pm$. }\label{table:massive}
\end{table}

\begin{table}[h!]
\centering
\begin{tabular}{|c|c|c|c|}
\hline
Bases & ~~~~Plane Waves~ ~~~
&~~~ Conformal Primary Wavefunctions~ ~~\\
\hline
&&\\
~~~\text{Notations}~~~&$~  \exp\left[ \pm i k\cdot X\right]~$  
&$\varphi_{\Delta }^\pm (X^\mu; \vec w) = \left(-q(\vec w)\cdot X\mp i\epsilon \right)^{-\Delta}$ \\
&&\\
\hline
&&\\
Labels & $k^\mu~(k^2=0,~k^0>0)$&$ \Delta\in\frac d2 +i \mathbb{R}
~,~ \vec w\in\mathbb{R}^d$\\
&&\\
\hline
\end{tabular}
\caption{A comparison between the plane wave basis  and the  conformal primary basis $\mathcal{B}^\pm_{m=0}$ for  normalizable, outgoing/incoming solutions to the massless Klein-Gordon equation. The plane wave is labeled by  a null momentum $k^\mu$, whereas the conformal primary wavefunction is labeled by $\Delta \in  \frac d2 + i\mathbb{R}$ and $\vec w\in\mathbb{R}^d$.  There is another basis $\widetilde {\mathcal{B}}_{m=0}^\pm$  that is shadow to $\mathcal{B}^\pm_{m=0}$.}\label{table:massless}
\end{table}

We  then discuss massless conformal primary wavefunctions with spin.   More specifically, we will construct solutions to the $(d+2)$-dimensional Maxwell and vacuum linearized Einstein equation that transform as $d$-dimensional spin-one and spin-two conformal primaries, respectively.  This extends the study of  spin-one conformal primary wavefunctions in $(3+1)$ dimensions  of  \cite{Cheung:2016iub}. 
 One qualitative difference between the massless spinning wavefunctions and the scalar wavefunctions is the presence of gauge or diffeomorphism symmetry. 
It turns out that  conformal covariance of the on-shell wavefunction selects a particular gauge.  We will also discuss spinning conformal primary wavefunctions that  are pure gauge/diffeomorphism in diverse spacetime dimensions.  
Finally we show that the  plane waves are spanned by  spinning conformal primary wavefunctions on the principal continuous series \eqref{introprincipal}.  
The transition from momentum space to the space of massless spinning conformal primary wavefunctions is implemented by a Mellin transform (or plus a shadow transform).

Scattering amplitudes written in the conformal primary basis manifestly enjoy the conformal covariance of  $d$-dimensional conformal correlators.  For example, the three-point decay amplitude  of a four-dimensional $\phi^3$ theory written in this basis was shown to take the form of a two-dimensional CFT three-point function in a special mass limit  \cite{Pasterski:2016qvg}.  The soft limit and collinear singularities of gluon and graviton amplitudes were analyzed in this basis in \cite{Cheung:2016iub}. The factorization singularity of  scattering amplitudes has also been studied in the CFT language  \cite{Cardona:2017keg,NVWZ}.  It will be interesting to explore how other conformal structures can be translated into statements about  scattering amplitudes in this basis.

The construction of conformal primary wavefunctions proceeds naturally via the embedding formalism in CFT \cite{Dirac:1936fq,Mack:1969rr,Cornalba:2009ax,Weinberg:2010fx,Costa:2011mg,Costa:2011dw,SimmonsDuffin:2012uy,Costa:2014kfa}.  Our flat space conformal wavefunctions are expressed in terms of the hyperbolic space $H_{d+1}$ bulk-to-boundary propagators lifted to the embedding Minkowski space.  
 In the CFT context, the embedding Minkowski space $\mathbb{R}^{1,d+1}$ is merely a fictitious space one introduces to realize the conformal transformation linearly.  By contrast, in the current setting the embedding $\mathbb{R}^{1,d+1}$  is the spacetime where physical scattering processes take place.

The rest of the paper is organized as follows. In Section \ref{sec:massive}, we review and extend the definition of massive scalar conformal primary wavefunctions in general spacetime dimensions.  In Section \ref{sec:basis}, we determine the range of the  conformal dimension to be the principal continuous series  of $SO(1,d+1)$.  In Section \ref{sec:massless}, we consider massless scalar conformal primary wavefunctions and determine the range of their conformal dimensions.  In particular, we  show that the change of basis from  momentum space to  conformal primary wavefunctions is implemented by a Mellin transform in the massless case.  In Sections \ref{sec:photon} and \ref{sec:graviton} we discuss massless spin-one and spin-two conformal primary wavefunctions, respectively.

\section{Massive Conformal Primary Wavefunctions}\label{sec:massive}

In this section we construct massive scalar wavefunctions in $(d+2)$-dimensional Minkowski spacetime $\mathbb{R}^{1,d+1}$ with coordinates $X^\mu$, $\mu=0,1,\cdots, d+1$.

\subsection{Massive Scalar Conformal Primary Wavefunctions in General Dimensions}

Let us review the massive scalar conformal primary wavefunction  defined in \cite{Pasterski:2016qvg}. The \textit{massive scalar conformal primary wavefunction} $\phi_\Delta(X^\mu;\vec w)$ of mass $m$ in $\mathbb{R}^{1,d+1}$ is a wavefunction labeled by a ``conformal dimension" $\Delta$ and a point $\vec w$ in $\mathbb{R}^d$.  It satisfies the  following two defining properties:\footnote{We omit the mass $m$ dependence of the conformal primary wavefunction in the notation $\phi_\Delta(X^\mu;\vec w)$.}
\begin{itemize}
\item It satisfies the $(d+2)$-dimensional  massive Klein-Gordon equation of mass $m$,\footnote{Our  convention for the spacetime signature in $\mathbb{R}^{1,d+1}$ is $(-++\cdots +)$.}
\begin{align}\label{KGsol}
\left( {\partial\over \partial X^\nu } {\partial\over \partial X_\nu} -m^2  \right) \phi_{\Delta}(X^\mu ;\vec w)=0\,.
\end{align}
\item It transforms covariantly as a scalar conformal primary operator in $d$ dimensions under an $SO(1,d+1)$ transformation,
\begin{align}\label{covariance}
\phi_\Delta \left( \Lambda^\mu_{~\nu} X^\nu ; \vec w\,'(\vec w) \right)
= \left| {\partial \vec w\,' \over\partial \vec w}\right|^{-\Delta/d} \,
\phi_\Delta( X^\mu ; \vec w)\,,
\end{align}
where $\vec w\,'(\vec w)$ is an $SO(1,d+1)$ transformation that acts non-linearly on $\vec w\in\mathbb{R}^d$ and $\Lambda^\mu_{~\nu}$ is the associated group element in the $(d+2)$-dimensional representation.  More explicitly, $\vec w\,'(\vec w)$ is generated by:
\begin{align}\label{CT}
&\mathbb{R}^d~\text{translation}:~~~~~~\vec w\,'  = \vec w+\vec a\,,\notag\\
&SO(d) ~\text{rotation}:~\,~~~ \vec w\,'= M\cdot \vec w\,,\\
&\text{dilation}:~~~~~\,~
~ ~~~~~~~\vec w\,'= \lambda \vec w\,, \notag\\
&\substack{\text{special conformal}\\\text{transformation}}:~~~~~~~~
\vec w\,'  ={\vec w + |\vec w|^2 \vec b \over
 1+2 \vec b \cdot \vec w +|\vec b|^2 |\vec w|^2}\,.  \notag
\end{align}
\end{itemize}

Being a solution to the Klein-Gordon equation, the conformal primary wavefunction can be expanded on the plane waves.       The Fourier expansion takes the form of an integral over all the possible outgoing or incoming on-shell momenta, each of which is a copy of the $(d+1)$-dimensional hyperbolic space $H_{d+1}$.  
To be more concrete, let $y,\vec z$ be the coordinates of $H_{d+1}$ with $y>0$ and $\vec z\in \mathbb{R}^d$.  The  $H_{d+1}$ metric  is
\begin{align}
ds^2 _{H_{d+1}} =  {dy^2 +d \vec z \cdot d\vec z\over y^2}\,,
\end{align}
with $y=0$ being the boundary.  This geometry has an $SO(1,d+1)$ isometry $\vec z \to \vec z\,' (y,\vec z)\,, y\to y' (y,\vec z)$  that is generated by
\begin{align}\label{isometry}
&\mathbb{R}^d~\text{translation}:~\,~~ y'  =y\,,~~~~~~~~\vec z\,'  = \vec z+\vec a\,,\notag\\
&SO(d) ~\text{rotation}:~ ~y' =y\,,~\,~~~~~~~ \vec z\,'= M\cdot \vec z\,,\\
&\text{dilation}:~~~~~~~~\,~~
~ y'= \lambda y\,,~~~~~~~\vec z\,'= \lambda \vec z\,, \notag\\
&\substack{\text{special conformal}\\\text{transformation}}:~~~~~
y'= {y\over 1+2 \vec b \cdot \vec z +|\vec b|^2 (y^2+|\vec z|^2)}\,,
~~~~
\vec z\,'  ={\vec z + (y^2 +|\vec z|^2) \vec b \over
 1+2 \vec b \cdot \vec z +|\vec b|^2 (y^2+|\vec z|^2)}\,.  \notag
\end{align} 

We can then parametrize a unit timelike vector $\hat p(y,\vec z)$  satisfying $\hat p^2=-1$ in terms of the $H_{d+1}$ coordinates as,
\begin{align}\label{phat}
\hat p (y,\vec z) = \left( {1+y^2 + |\vec z|^2 \over 2y} \,,  \, { \vec z \over y} \,, \,{1-y^2 - |\vec z|^2 \over 2y}    \right)\,.
\end{align}
The map $\hat p(y,\vec z)$ defines an embedding of the $H_{d+1}$ into the upper branch ($\hat p^0>0$) of the unit hyperboloid in $\mathbb{R}^{1,d+1}$.  We will henceforth use $\hat p$ and $(y,\vec z)$ interchangeably to parametrize a point in $H_{d+1}$.  The advantage of working with $\hat p^\mu$ is that the non-linear $SO(1,d+1)$ action \eqref{isometry} on $y,\vec z$ now becomes linear on $\hat p^\mu$,
\begin{align}\label{ptransform}
\hat p^\mu ( y' ,\vec z\,' )  = \Lambda^\mu_{~\nu} \hat p^\nu\,,
\end{align}
where $\Lambda^\mu_{~\nu}$ is the associated group element of $SO(1,d+1)$ in the $(d+2)$-dimensional representation.

One last ingredient we need is the scalar bulk-to-boundary propagator $G_\Delta(\hat p;\vec w)$ in $H_{d+1}$ \cite{Witten:1998qj},
\begin{align}\label{G}
G_{\Delta}(\hat p; \vec w)  = \left( {y\over y^2+ |\vec z- \vec w|^2}\right)^\Delta\,,
\end{align}
where $\vec w \in \mathbb{R}^d$ is a  point on the boundary of $H_{d+1}$.  Let us define a map from $\mathbb{R}^d$ to a ``unit" null momentum $q^\mu$ in $\mathbb{R}^{1,d+1}$ as
\begin{align}\label{qmap}
q^\mu (\vec w) =   \left( \, 1+ |\vec w|^2 \,,\,  2\vec w \, , \, 1-|\vec w|^2  \, \right)\,.
\end{align}
We will use $\vec w$ and $q^\mu$ interchangeably to parametrize a point in $\mathbb{R}^d$.  While $\vec w$ transforms non-linearly under $SO(1,d+1)$ as in \eqref{CT}, its embedding $q^\mu$ into $\mathbb{R}^{1,d+1}$ transforms linearly,
\begin{align}\label{qtransform}
q^\mu (\vec w\,')  = \left| {\partial \vec w\,' \over \partial \vec w}\right|^{1/d} \, \Lambda^\mu_{~\nu} q^\nu(\vec w) \,.
\end{align}
We will often use 
\begin{align}\label{daq}
\p_a q^\mu \equiv {\p \over \p w^a }q^\mu(\vec w)  
= 2( w^a \,,\, \delta^{ba} \,,\, -w^a)\,.
\end{align}
In terms of the coordinates $\hat p^\mu(y,\vec z)$ and $q^\mu(\vec w)$, the  bulk-to-boundary propagator can be written succinctly as \cite{Costa:2014kfa}
\begin{align}\label{Gembedding}
G_\Delta( \hat p ;q )  = {1\over (-\hat p \cdot q)^\Delta}\,.
\end{align}
Under an $SO(1,d+1)$ transformation  $\hat p(y,\vec z)\to \hat p' = \hat p(y',\vec z\,')$ and $q(\vec w) \to q' = q(\vec w\,')$, the $G_\Delta$ transforms covariantly as
\begin{align}\label{Gcovariance}
G_\Delta( \hat p '; q') =  \left| {\partial \vec w\,' \over \partial \vec w}\right|^{-\Delta/d}  \,G_\Delta(\hat p; q)\,.
\end{align}

With the above preparation, we can now write down  the Fourier expansion of the scalar conformal primary wavefunction on the plane waves,
\begin{align}\label{CPW}
\boxed{\,
\phi^\pm_\Delta(X^\mu ;  \vec w) 
= \int_{H_{d+1}} [d\hat p ]  \, G_\Delta ( \hat p ;\vec w)  \, \exp \left[
\, \pm i m \hat p \cdot X \,\right]
\,}\,,
\end{align}
with Fourier coefficients being the scalar bulk-to-boundary propagator in $H_{d+1}$.  
We use a  plus (minus) sign  for an outgoing (incoming) wavefunction.  Here $[d\hat p]$ is the $SO(1,d+1)$ invariant measure on $H_{d+1}$:
\begin{align}
\int_{H_{d+1}} [d \hat p] \equiv \int_0^\infty {dy\over y^{d+1} } \int d^d\vec z  =  \int {d^{d+1} \hat p^i \over \hat p^0 } \,,
\end{align}
where $i=1,2,\cdots,d+1$ and $\hat p^0 = \sqrt{ \hat p^i \hat p^i +1}$. 
The conformal primary wavefunction given in \eqref{CPW} satisfies the defining property \eqref{covariance} thanks to the conformal covariance of the bulk-to-boundary propagator \eqref{Gcovariance}.

Importantly, the conformal dimension $\Delta$ of the conformal primary wavefunction $\phi^\pm_\Delta(X^\mu;\vec w)$ is \textit{not} related to the mass $m$.  Indeed, $\Delta$ together with $\vec w\in\mathbb{R}^d$ should be thought of as the dual variables to an on-shell momentum $p^\mu$ that label the space of solutions to the Klein-Gordon equation.  In particular, we require a continuum of conformal primary wavefunctions $\phi_\Delta (X^\mu ; \vec w)$ to form a basis of normalizable wavefunctions.  We will determine the range of $\Delta$ in Section \ref{sec:basis}.

\subsection{Closed-Form Expression}

In \eqref{CPW} we have provided an integral representation for the massive scalar conformal primary wavefunction.  In this section we will write its closed-form expression in terms of Bessel functions by directly solving the Klein-Gordon equation.

Let us consider the following ansatz for the wavefunction:
\begin{align}\label{ansatz}
\phi_\Delta (X^\mu ;\vec w ) =  {f(X^2) \over (-q\cdot X)^\Delta}\,,
\end{align}
with $q^\mu = q^\mu(\vec w)$ given in \eqref{qmap}.  The factor $1/(-q\cdot X)^\Delta$ solves the massless Klein-Gordon equation and has the desired conformal covariance \eqref{covariance} following from \eqref{qtransform},\footnote{This implies that $1/(-q\cdot X)^\Delta$ is a \textit{massless} conformal primary wavefunction, which will be discussed  in full detail in Section \ref{sec:massless}.} while the numerator is invariant under $SO(1,d+1)$.  Hence the ansatz \eqref{ansatz} already obeys \eqref{covariance} and we only need to solve for $f(X^2)$ such that $\phi_\Delta$ is a solution to the massive Klein-Gordon equation.  The massive Klein-Gordon equation gives the following differential equation for $f(X^2)$:
\begin{align}
0= 4X^2 f''(X^2) -2( 2\Delta-d-2) f'(X^2) -m^2 f(X^2)\,,
\end{align}
from which we obtain
\be\label{f}
f(X^2)=(\sqrt{-X^2})^{\Delta-\frac d2}\left[c_1\, I_{\Delta-\frac d2}(m\sqrt{X^2})+c_2\, I_{-\Delta+\frac d2}(m\sqrt{X^2})\right]
\ee
where $I_{\alpha}(x)$ is the modified Bessel function of the first kind.  For large spacelike $X^\mu$, the Bessel function $I_{\alpha}(m\sqrt {X^2})$ grows exponentially as $e^{m\sqrt{X^2}}$, so a generic solution \eqref{f} will not give rise to normalizable wavefunctions.  
   By requiring  a finite Klein-Gordon norm \eqref{KGip} of the wavefunction, we select out a particular linear combination that is proportional to a modified Bessel of the second kind, $K_\alpha(x)=\frac{\pi}{2}\frac{I_{-\alpha}(x)-I_{\alpha}(x)}{\sin(\alpha\pi)}$, which dies off exponentially as $e^{- m\sqrt{X^2}}$ for large $X^2$. We can fix the overall constant by comparing with the integral expression \eqref{CPW} and find\footnote{To obtain the above closed-form expression, we should analytically continue to imaginary $m$  as in \cite{Pasterski:2016qvg} and perform the integral.}
\be\label{closedform}
\phi_{\Delta}^\pm(X^\mu;\vec{w})= 
{ 2^{\frac d2+1} \pi^{\frac d2} \over (im)^{d\over2}}
{(\sqrt{-X^2})^{\Delta-\frac d2}  \over (-q(\vec w)\cdot X\mp i \epsilon)^{\Delta}}
K_{\Delta-\frac d2}(m\sqrt{X^2})\,.
\ee
We have introduced an $i\epsilon$ prescription for the denominator.

\subsection{Shadow Transform}\label{sec:shadow}

In this section we show that the conformal primary wavefunction $\phi_{\Delta}^\pm$ is the shadow transform of $\phi_{d-\Delta}^\pm $.\footnote{We would like to thank Andy Strominger for pointing this out to us.}

Given a $d$-dimensional scalar conformal primary operator $\mathcal{O}_\Delta(\vec w)$, its \textit{shadow} \cite{Ferrara:1972xe,Ferrara:1972ay,Ferrara:1972uq,Ferrara:1973vz} $\mathcal{\widetilde  O}_\Delta(\vec w)$ is a non-local operator defined as
\begin{align}\label{scalarshadow}
\mathcal{\widetilde  O}_\Delta (\vec w) \equiv  
{\Gamma(  \Delta) \over \pi^{\frac d2}\Gamma(\Delta-\frac d2)}
\int d^d\vec w\,' {1\over |\vec w-\vec w\,'|^{2(d-\Delta)} }\,\mathcal{O}_\Delta(\vec w \,')\,.
\end{align}
The shadow operator $\mathcal{\widetilde O}_\Delta$ transforms as a scalar conformal primary operator with conformal dimension $d-\Delta$.  The normalization constant is chosen so that our conformal primary wavefunctions transform nicely under 
\eqref{scalarshadow}.


More generally, given a $d$-dimensional  conformal primary $\mathcal{O}_{a_1\cdots a_J}(\vec w)$ in the symmetric traceless rank-$J$ representation of $SO(d)$ with dimension $\Delta$, its shadow $\widetilde {\mathcal{O}}_{a_1\cdots a_J}(\vec w)$ can be most conveniently computed in terms of its uplift  $\mathcal{O}_{\mu_1\cdots \mu_J}(\vec w)$ to the embedding  space $\mathbb{R}^{1,d+1}$ \cite{SimmonsDuffin:2012uy}:
\begin{align}\label{spinshadow}
\widetilde {\mathcal{O}}_{\mu_1 \cdots \mu_J}(\vec w)
=
{\Gamma(\Delta+J) \over \pi^{d\over2} (\Delta-1)_J\Gamma(\Delta-\frac d2)}
 \int d^d \vec w\,' \, 
{\prod_{n=1}^J
 \left[ \, \delta^{\nu_n}_{\mu_n} (-\frac 12q\cdot q')
 +\frac 12{q'}_{\mu_n} q^{\nu_n} \,\right]
 \over (-\frac 12 q\cdot q')^{d-\Delta+J}
 }
 \mathcal{O}_{\nu_1\cdots \nu_J} (\vec w\,')
 \,,
\end{align}
where $(a)_J \equiv \Gamma(a+J)/\Gamma(a)$ and $q^\mu = q^\mu(\vec w)$ as in \eqref{qmap}.  The uplifted operator $\mathcal{O}_{\mu_1\cdots \mu_J}(\vec w)$ is transverse to $q^\mu$ and is defined modulo terms of the form $q^{\mu_i} \Lambda^{\mu_1 \cdots \hat \mu_i \cdots \mu_J}(\vec w)$.  Note that $-\frac 12 q\cdot q' = |\vec w-\vec w\,'|^2$. We recover the $d$-dimensional primary $\mathcal{O}_{a_1\cdots a_J}(\vec w)$ via the projection:
\begin{align}\label{projection}
\mathcal{O}_{a_1\cdots a_J}(\vec w)
= {\p q^{\mu_1} \over\p w^{a_1} }\cdots  {\p q^{\mu_J} \over\p w^{a_J} } \mathcal{O}_{\mu_1\cdots \mu_J}(\vec w)\,,
\end{align}
and similarly for its shadow $\widetilde {\mathcal{O}}_{a_1\cdots a_J}(\vec w)$.  
The shadow operator $\widetilde {\mathcal{O}}_{a_1\cdots a_J}(\vec w)$ transforms as a  spin-$J$ conformal primary with dimension $d-\Delta$ under $SO(1,d+1)$.

To study the shadow transform of our conformal primary wavefunction, let us first note a useful identity  (see, for example, \cite{SimmonsDuffin:2012uy})
\begin{align}\label{david}
\int d^d \vec z {1\over |\vec z-\vec w|^{2(d-\Delta)} } {1\over (-q(\vec z)\cdot X)^\Delta}
={\pi^{d\over2}  \Gamma(\Delta- \frac d2) \over \Gamma(\Delta)}
{(-X^2)^{\frac d2 -\Delta} \over (-q(\vec w) \cdot X)^{d-\Delta}}\,.
\end{align}
Now the shadow transform of our conformal primary wavefunction \eqref{closedform} directly follows from  \eqref{david}:
\begin{align}\label{shadowCPW}
\boxed{\,
\widetilde{\phi^\pm_{\Delta}}(X; \vec w) = \phi^\pm_{d-\Delta} (X; \vec w) 
 \, }\,.
\end{align}
Thus the conformal primary wavefunctions $\phi^\pm_\Delta$  and $\phi^\pm_{d-\Delta}$ should not be counted as linearly independent solutions to the massive Klein-Gordon equation as they are related by a shadow transform.

\subsection{Integral Transform: From Amplitudes to Conformal Correlators}

In \eqref{CPW} we determined the change of basis from the plane wave $e^{\pm i p\cdot X}$ to  $\phi_\Delta^\pm(X^\mu;\vec w)$ for a single wavefunction. This change of basis can be imminently extended to any $n$-point scattering amplitude  of massive scalars.  Let $\mathcal{A}(p_i^\mu)$ be such an amplitude in momentum space, including the momentum conservation delta function $\delta^{(d+2)}(\sum_i p_i^\mu)$.  We can then define an integral transform that takes this amplitude to the basis of conformal primary wavefunctions,
\begin{align}\label{integral}
\mathcal{\widetilde A}(\Delta_i , \vec w_i )  
\equiv \prod_{k=1}^n \int_{H_{d+1}} [d\hat p_k] \, 
G_{\Delta_k} (\hat p _k ; \vec w_k )  \, \
\mathcal{A} ( \pm m_i \hat p_i^\mu)\,,
\end{align}
where we have parametrized an outgoing (incoming) on-shell momentum as $p^\mu _i  =m_i \hat p^\mu$ ($p^\mu _i  = -m_i \hat p^\mu$) with $\hat p_i^2=-1$ as in \eqref{phat}. 
Thanks to  the conformal covariance \eqref{covariance} of the  wavefunctions $\phi^\pm_\Delta(\vec w)$, the scattering amplitude in this basis transforms covariantly as a $d$-dimensional CFT $n$-point function of scalar primaries with dimensions $\Delta_i$,
\begin{align}\label{conformalcorrelator}
\mathcal{\widetilde A}(\Delta_i, \vec w_i'(\vec w_i) ) 
= \prod_{k=1}^n \left|  {\partial \vec w_k' \over \partial \vec w_k}\right|^{-\Delta_k/d}\,
\mathcal{\widetilde A}(\Delta_i,\vec w_i ) \,. 
\end{align}
Hence the change of basis \eqref{integral} is implemented as an integral transformation that takes a scattering amplitude $\mathcal{A}(p_i^\mu)$ to a $d$-dimensional conformal correlator $\mathcal{\widetilde A}(\Delta_i,\vec w_i )$.

\section{A Conformal Primary Basis}\label{sec:basis}

It is natural to ask for what, if any, range of the conformal dimension $\Delta$ will the set of conformal primary wavefunctions $\phi^\pm_\Delta(X^\mu; \vec w)$ form a  basis for delta-function-normalizable, outgoing/incoming solutions of the Klein-Gordon equation.\footnote{Throughout this paper, we will consider the space of \textit{complex} solutions to the wave equation.  As usual, a reality condition is needed if one wants to perform a mode expansion of a real field on these complex solutions.}  In this section we show that the range of $\Delta$ can be chosen to be the principal continuous series  of $SO(1,d+1)$.

\subsection{Inverse Transform and Principal Continuous Series}

We begin by seeking the inverse transform of \eqref{CPW}, i.e. the expansion of plane waves into the conformal primary wavefunctions.\footnote{We would like to thank H.-Y. Chen, X. Dong,  J. Maldacena, and  H. Ooguri for discussions on this point.}  Since the plane waves form a basis, if we can expand them on a certain set of conformal primary wavefunctions, then the latter also forms a (possibly over-complete) basis.   This is possible if the bulk-to-boundary propagator, which is the Fourier coefficient in \eqref{CPW}, satisfies certain orthonormality conditions for some range of the conformal dimension $\Delta$.  Indeed, this is the case if $\Delta$ belongs to the  \textit{principal continuous series} of the irreducible unitary $SO(1,d+1)$ representations,\footnote{More precisely, the principal continuous series representations are labeled by a conformal dimension $\Delta\in\frac d2 +i\mathbb{R}$ and  a representation, the spin, of $SO(d)$.  In this section we will consider the spin-zero principal continuous series representations, while in Section \ref{sec:photon} and \ref{sec:graviton} we will encounter spin-one and spin-two (i.e. symmetric traceless rank-two tensors of $SO(d)$) representations, respectively.}
\begin{align}
\boxed{\, 
\Delta \in \frac d2 +i \mathbb{R}\, }\,.
\end{align}
One orthonormality condition we need for  the $H_{d+1}$ scalar bulk-to-boundary propagator is \cite{Costa:2014kfa}:
 \begin{align}\label{ortho1}
 \int_{-\infty}^\infty d\nu \,
\mu(\nu)
 \,  \int d^d\vec w \, G_{\frac d2+i\nu}(\hat p_1; \vec w)G_{\frac d2-i\nu}(\hat p_2;\vec w)   =
   \, \delta^{(d+1)}(\hat p_1,\hat p_2)\,,
 \end{align}
 where $\delta^{(d+1)}(\hat p_1,\hat p_2)$ is the $SO(1,d+1)$ invariant delta function in $H_{d+1}$.  The measure factor $\mu(\nu)$ is
 \begin{align}
 \mu(\nu)   = {\Gamma(\frac d2 +i\nu  )\Gamma(\frac d2 -i\nu) \over
  4\pi^{d+1}  \Gamma(i\nu)\Gamma(-i\nu)}\,,
 \end{align}
which is an even, non-negative function of $\nu$. 
The second orthonormality condition is \cite{Costa:2014kfa}:
 \begin{align}\label{ortho2}
& \int_{H_{d+1}}[d\hat p]\, G_{\frac d2+i\nu}(\hat p;\vec w_1) 
 G_{\frac d 2+i \bar\nu}(\hat p;\vec w_2)
 =\\
&2\pi^{d+1} 
{\Gamma(i\nu)\Gamma(-i\nu) \over \Gamma(\frac d2 +i\nu) \Gamma(\frac d2 -i\nu)}
 \delta(\nu+\bar \nu) 
 \delta^{(d)}(\vec w_1-\vec w_2)
 + 2\pi^{\frac d2+1} 
   {\Gamma(i\nu)\over \Gamma(\frac d2+i\nu)}
     {\delta(\nu-\bar\nu)}
   {1\over |\vec w_1-\vec w_2|^{2(\frac d2+i\nu)}}\,.\notag
 \end{align}

Now we are ready to write down the inverse transform of \eqref{CPW}.  Combining \eqref{CPW} and \eqref{ortho1}, we immediately obtain:
\begin{align}\label{inverse}
\boxed{\, 
e^{\pm im \hat p\cdot X}  =2\int_{0}^\infty d\nu \,\mu(\nu) \int d^d\vec w \,\,
G_{\frac d2-i\nu}(\hat p;\vec w) \,\, \phi^\pm_{\frac d2+i\nu}(X^\mu;\vec w)  \,}\,,
\end{align}
where we have used  \eqref{david} and \eqref{shadowCPW} to rewrite the expansion  only on wavefunctions with non-negative $\nu$.

Given that the plane waves form a  basis for the normalizable solutions of the Klein-Gordon equation, it is tempting to conclude from \eqref{inverse} that the conformal primary wavefunctions on the principal continuous series with non-negative $\nu$ form a  basis  too.  We have to check the following two conditions, however, in order to  prove the  above assertion:
\begin{itemize}
\item Are  the conformal primary wavefunctions $\phi^\pm_{\Delta}(X^\mu ; \vec w)$ with $\Delta \in \frac d2 +i \mathbb{R}_{\ge0}$ and $\vec w\in \mathbb{R}^d$  linearly independent of each other?\\
\item Are the conformal primary wavefunctions  delta-function-normalizable with respect to the Klein-Gordon norm?
\end{itemize}
  We will shortly give   positive answers to   both questions in Section \ref{sec:2pt} by explicit computation of the Klein-Gordon inner product between conformal primary wavefunctions.

\subsection{Klein-Gordon Inner Product}\label{sec:2pt}

In studying the solution space for the massive Klein-Gordon equation, we focus on wavefunctions that are (delta-function-)normalizable with respect to a certain inner product.  A natural inner product between complex wavefunctions is the Klein-Gordon inner product defined as
\begin{align}\label{KGip}
( \Phi_1 , \Phi_2)&=- i \int d^{d+1}X^i~\left[\,  
\Phi_{1} ( X) \, \partial_{X^0}\Phi_{2}^{*} ( X)
-\partial_{X^0}\Phi_{1} ( X) \,\Phi_{2}^{*} ( X)
\right] \,,
\end{align}
where $i=1,2,\cdots, d+1$ is an index for the spatial directions in $\mathbb{R}^{1,d+1}$ and $*$ stands for complex conjugation. 
 Using the Klein-Gordon equation, one can show that the above inner product does not depend on the choice of the Cauchy surface we integrate over.
The plane waves, for example, are delta-function-normalizable with respect to this inner product:\footnote{Strictly speaking,  while \eqref{KGip} is  a positive-definite inner product on the space of outgoing (i.e. positive energy) wavefunctions, we should use the minus of \eqref{KGip} as a positive-definite inner product on the space of incoming (i.e. negative energy) wavefunctions.}
\begin{align}\label{planewaveip}
( e^{\pm i p\cdot X}  , e^{\pm i p' \cdot X} ) 
= \pm 2(2\pi)^{d+1} \, p^0 \, \delta^{(d+1)} (p^i - {p^i}'  )\,.
\end{align}
Furthermore, they form a basis of normalizable solutions to the Klein-Gordon equation.

Let us compute the Klein-Gordon inner product of two conformal primary wavefunctions:
\begin{align}{\label{2pt}}
&\left( \phi^\pm_{\frac d2+i\nu_1} (X^\mu ;\vec w_1), \phi^\pm_{\frac d2+i\nu_2} (X^\mu ;\vec w_2) \right)
 \notag\\
 &~~~~~~~~~~~~~~~~~~= 
 \pm{ 2^{d+3} \pi^{2d+2} \over m^{d} }
 {\Gamma(i\nu_1) \Gamma(-i\nu_1) \over \Gamma(\frac d2 +i\nu_1)
 \Gamma(\frac d2-i\nu_1)}\,
 \delta(\nu_1- \nu_2) \, \delta^{(d)}(\vec w_1- \vec w_2)\notag\\
& ~~~~~\,~~~~~~~~~~~~~~~~ \pm  {2^{d+3} \pi^{{3d\over 2} +2}\over m^d}
 {\Gamma(i\nu_1) \over \Gamma(\frac d2+i\nu_1)} 
 \delta(\nu_1+\nu_2)
 {1\over |\vec w_1-\vec w_2|^{2(\frac d2+ i\nu_1)}}\,,
 \end{align}
where we have used the second orthonormality condition \eqref{ortho2}.  As we saw in the last section, conformal primary wavefunctions with negative $\nu$ are linearly related to those with positive $\nu$ by the shadow transform.  Therefore we only need to consider the inner product between those with positive $\nu$, in which case the second term in \eqref{2pt} drops out.  Note that the coefficient of the surviving term  (i.e. the first term) is positive (negative) definite for outgoing (incoming) wavefunctions, as it should be for an inner product on single particle solutions.

From \eqref{2pt}, we then conclude that the conformal primary wavefunctions are delta-function-normalizable with respect to the Klein-Gordon inner product. Furthermore, the set of conformal primary wavefunctions $\phi^\pm_{\frac d2 +i\nu}(X^\mu;\vec w)$ with non-negative $\nu$ and $\vec w\in\mathbb{R}^d$ are orthogonal to each other, and hence are linearly independent.

\subsection{The Massive Scalar Conformal Primary  Basis}\label{sec:massivebasis}

We are finally ready to write down the complete set of linearly independent conformal primary wavefunctions that span the  space of  outgoing or incoming solutions to the massive Klein-Gordon equation.  We will call such  bases of wavefunctions the outgoing ($+$) and incoming $(-)$ \textit{conformal primary bases} $\mathcal{B}^\pm$.

Let us recap the logic.  In \eqref{inverse} we showed that the plane waves can be expanded upon conformal primary wavefunctions on the principal continuous series  with non-negative $\nu$, so the latter must span the whole solution space.  
In \eqref{2pt} we showed that the conformal primary wavefunctions with non-negative $\nu$ and $\vec w\in\mathbb{R}^d$ are delta-function-normalizable and linearly independent of each other.  Thus we conclude that the  outgoing/incoming conformal primary  bases $\mathcal{B}^\pm$ for the massive Klein-Gordon equation can be chosen to be
\begin{align}\label{massivebasis}
\boxed{
\mathcal{B}^\pm=  \left\{\,  \phi^\pm_{\frac d2 + i\nu} (X; \vec w) \,  \Big|\,  \nu\ge0, \vec w\in \mathbb{R}^d \,\right \}\, }\,,
\end{align}
where we recall that the plus and minus superscripts denote outgoing and incoming wavefunctions, respectively.  We contrast the massive conformal primary bases with the plane wave bases in Table \ref{table:massive}.

Alternatively, the shadows of the bases \eqref{massivebasis} are equally good conformal primary bases for the massive Klein-Gordon solutions:
\begin{align}\label{massiveshadowbasis}
\boxed{
\mathcal{\widetilde B}^\pm=  \left\{\,  \phi^\pm_{\frac d2 + i\nu} (X; \vec w) \,  \Big|\,  \nu\le0, \vec w\in \mathbb{R}^d \,\right \}\, }\,.
\end{align}
To sum up, we have identified a pair of bases $\mathcal{B}^+$ and $\mathcal{\widetilde B}^+$ for the outgoing, normalizable solutions to the massive Klein-Gordon equation. Similarly, we have identified a pair of bases $\mathcal{B}^-$ and $\mathcal{\widetilde B}^-$ for the incoming, normalizable solutions to the massive Klein-Gordon equation.  The four bases are related by complex conjugation and shadow transformation as below:
\begin{align}
\begin{split}
&\text{Outgoing}:~~~~\mathcal{B}^+   ~~~~ \underset{\text{shadow}}{\longleftrightarrow}~~~~ \mathcal{\widetilde B}^+\\
&~~~~~~~~~~~~~~\,~~~~~\Big\updownarrow_{\text{c.c}} ~~~~~~~~~~~~~~~\Big\updownarrow_{\text{c.c}} \\
&\text{Incoming}:~~~~\mathcal{\widetilde B}^-   ~~~~ \underset{\text{shadow}}{\longleftrightarrow}~~~~  \mathcal{B}^-
\end{split}
\end{align}

For odd $d$, there are also principal discrete series irreducible unitary $SO(1,d+1)$ representations with conformal dimension $\Delta= \frac d2 +\mathbb{Z}_+$. The conformal primary wavefunctions for the discrete series, however, are \textit{not} normalizable with respect to the Klein-Gordon inner product \eqref{KGip}.  To see this, note that the Klein-Gordon inner product between two conformal primary wavefunctions with conformal dimensions $\Delta_1$ and $\Delta_2$ is proportional to 
\begin{align}
\int_{H_{d+1}} [d\hat p] \, 
 G_{\Delta_1} (\hat p ;\vec w_1) 
 \, G_{\Delta_2^*} (\hat p ;\vec w_2) \,,
\end{align}
which diverges for positive $\Delta_1$ and $\Delta_2$.  On the other hand, the wavefunction is delta-function-normalizable if $\Delta$ is on the principal continuous series.  Therefore we will not  consider the discrete series representations in this paper.

Recently, a simple  solution to the conformal crossing equation of $SO(1,d+1)$ was found \cite{Gadde:2017sjg} for all $d$.\footnote{See also \cite{Hogervorst:2017sfd} for a crossing solution in one dimension on the principal continuous series.} The spectrum of conformal primaries consists of the whole continuum of principal continuous series representations plus the principal discrete series.  It will be interesting to explore the connection between this conformal crossing solution and   scattering amplitudes in $\mathbb{R}^{1,d+1}$.

\section{Massless Scalars}\label{sec:massless}

So far we have been studying the conformal primary basis of the massive Klein-Gordon equation.  In this section we show how the massless limit of our massive conformal primary wavefunction reduces to a combination of the Mellin  transform of the plane wave and its shadow.   We will then determine the conformal primary basis of solutions to the massless Klein-Gordon equation with nonvanishing inner product.

\subsection{Mellin Transform}

Let us consider the massless limit of the massive conformal primary wavefunction written in the integral representation  \eqref{CPW}.  It will be convenient to  change the integration variable from $y$ to
\begin{align}
\omega \equiv {m \over 2y}\,.
\end{align}
The massless limit is taken by sending $y = {m\over 2\omega}\to0$ (while holding $\omega$ fixed) in the bulk-to-boundary propagator, 
\begin{align}\label{bdylimit}
G_\Delta(y,\vec z;\vec w)
~\underset{m\to0}{\longrightarrow }~
\pi^{d\over2}   {\Gamma(\Delta-\frac d2)\over\Gamma(\Delta)} 
y^{d-\Delta} \delta^{(d)}(\vec z-\vec w)
+{y^\Delta\over |\vec z-\vec w|^{2\Delta}}+\cdots\,,
\end{align}
where $\cdots$ are higher order terms in the small $y$ expansion.  Note that for our purpose $\Delta=\frac d2 + i \nu$,\footnote{Strictly speaking we have not justified that the massless conformal primary wavefunctions on the principal continuous series form a basis of wavefunctions.  We will show this in Section \ref{sec:m0basis}.} so the second term above is not smaller than the first one.

  The massless limit of the conformal primary wavefunction is \begin{align}\label{m0limit}
&\phi^\pm_{\frac d2 + i \nu } (X; \vec w) \,
\underset{m\to 0 }{\longrightarrow} \,
\left( {m\over2}\right)^{-\frac d2 - i\nu}
{\pi^{d\over2} \Gamma(i \nu) \over \Gamma(\frac d2 + i \nu)}
\int_0^\infty d\omega \, \omega^{\frac d2+i\nu - 1} e^{\pm i\omega q(\vec w)\cdot X}\notag\\
&\hspace{2.8cm}
+\left( {m\over2}\right)^{-\frac d2 +i\nu}
\int d^d \vec z {1\over |\vec z-\vec w|^{2(\frac d2+i\nu)}}
\int_0^\infty d\omega \omega^{\frac d2- i\nu-1}\,
e^{\pm i \omega q(\vec z) \cdot X}\,,
\end{align}
where $q^\mu(\vec w) = (1+|\vec w|^2 , 2\vec w , 1-|\vec w|^2)$ as in \eqref{qmap}.  We see that the massive conformal primary wavefunction does not have a well-defined massless limit because of the phases $m^{\pm i\nu}$. Nonetheless, we can extract the massless scalar conformal primary wavefunction from the coefficients of these phases.

The first term in \eqref{m0limit} takes the form of a Mellin transform of the plane wave.  This is the  \textit{massless scalar conformal primary wavefunction}, which, up to an overall constant, can be regularized as \cite{deBoer:2003vf,Campiglia:2015lxa,Cheung:2016iub,Pasterski:2016qvg,Campiglia:2017dpg}
\begin{align}\label{m0CPW}
\boxed{\, 
 \varphi^\pm _{\Delta} (X^\mu; \vec w)\equiv
\int_0^\infty d\omega \,\omega^{\Delta-1 }\,
e^{\pm i \omega q \cdot X-\epsilon \omega}
={ (\mp i)^\Delta \Gamma(\Delta)  \over  (-q(\vec w)\cdot X\mp i\epsilon)^{\Delta}}  \,}\,,
\end{align}
with $\epsilon>0$. 
Since we obtain $\varphi^\pm_\Delta(X^\mu;\vec w)$ from the massless limit of the conformal primary wavefunction, it automatically satisfies the defining properties \eqref{KGsol} (with $m=0$) and \eqref{covariance}.  The second term in \eqref{m0limit} is the shadow of $\varphi_\Delta^\pm(X^\mu;\vec w)$, and so is not a linearly independent wavefunction.  

In fact, the massless scalar conformal primary wavefunction \eqref{m0CPW} is, up to a normalization constant, nothing but the $H_{d+1}$ bulk-to-boundary propagator $G_\Delta( \hat p;\vec w)$  \eqref{G},   with the unit timelike vector $\hat p$ extended to a generic point $X^\mu$ in $\mathbb{R}^{1,d+1}$ \cite{Cheung:2016iub}.  Indeed, the bulk-to-boundary propagator satisfies the two defining properties \eqref{KGsol} and \eqref{covariance} of the conformal primary wavefunction. First, from \eqref{Gcovariance}, it manifestly has the desired conformal covariance \eqref{covariance}. Second, when extended to a generic point in $\mathbb{R}^{1,d+1}$,  it satisfies the massless Klein-Gordon equation via a hyperbolic slicing of the d'Alembert operator in flat space \cite{Cheung:2016iub,Campiglia:2015qka,Campiglia:2015kxa,Campiglia:2015lxa}. 
 We will make use of this fact for the spinning conformal primary wavefunctions in later sections.

Let us elaborate more on the properties of the massless conformal primary wavefunction.  For a fixed  null momentum $q^\mu$, the Lorentz boost along the spatial direction of $q^\mu$ acts as
\begin{align}
\text{Boost}:~q\cdot X \to \lambda \,( q\cdot X) ~
\Rightarrow ~ \varphi_\Delta(X^\mu ;\vec w) \to \lambda^{-\Delta} \varphi_\Delta(X^\mu;\vec w)\,,
\end{align}
where we have used \eqref{qtransform}. 
  Hence  the dilation of the conformal primary wavefunction $\varphi_\Delta^\pm$ is nothing but  a Lorentz boost  in $\mathbb{R}^{1,d+1}$.  
  
  The change of basis from  plane waves to  conformal primary wavefunctions is physically more intuitive in the massless case.  Given a null momentum $q^\mu$ in $\mathbb{R}^{1,d+1}$, it can be parametrized by a scale $\omega$ and a point $\vec w\in\mathbb{R}^d$   as in \eqref{qmap}.  The change of basis is implemented by a Mellin transform on this scale $\omega$, while the point $\vec w\in\mathbb{R}^d$ is directly identified as the position of the $d$-dimension conformal primary. The conformal dimension $\Delta$ is the dual variable for the scale $\omega$.  By contrast, there is no direct way to relate a timelike momentum $\hat p^\mu$ to a point in $\mathbb{R}^d$ in the massive case. The change of basis is done by a Fourier transform integrating over all $\hat p$ to conformal primary wavefunctions labeled by the dual variables $\Delta,\vec w$ as in \eqref{CPW}.

\subsection{The Massless Scalar Conformal Primary Basis}\label{sec:m0basis}

Since the massive conformal primary wavefunction does not have a well-defined massless limit, we should study the completeness question separately for the massless case.

Before we dive into the space of conformal primary wavefunctions, let us note a qualitative difference between the massless and massive solution spaces to the Klein-Gordon equation. In the massless case,  the constant wavefunction is a solution to the massless Klein-Gordon equation, which sits at the intersection between the outgoing and the incoming solution spaces.  On the other hand, the massive outgoing and incoming solution spaces are disjoint from each other.  The constant wavefunction has strictly zero-energy  and thus vanishing Klein-Gordon norm  \eqref{planewaveip}. \textit{We will exclude the constant wavefunction from our definition of either the outgoing or the incoming solution space}.  In fact, we will see that the conformal primary wavefunctions do not cover the constant wavefunction.

Let us start with the inverse transform of \eqref{m0CPW}. The inverse Mellin transform of the plane wave is (see, for example, \cite{deBoer:2003vf}),
\begin{align}
\,e^{\pm i \omega q\cdot X -   \epsilon \omega} =
\int_{-\infty}^\infty {d\nu\over 2 \pi} \,
 \omega^{-c -i\nu}
 {(\mp i )^{c+i\nu} \Gamma(c+i\nu) \over (-q\cdot X\mp i\epsilon )^{c +i\nu} }\,,~~~~~\omega>0\,,
\end{align}
where $c$ can be any positive number.  Hence all the massless plane waves \textit{except for the constant wavefunction} can be expanded on the conformal primary wavefunctions.  In other words, for any positive $c$, the massless conformal primary wavefunctions \eqref{m0CPW} with $\Delta\in c+i\mathbb{R}$ form a (possibly non-normalizable) basis of nonzero energy solutions to massless Klein-Gordon equations.

Next we need to determine for what value of $c >0$ will the massless conformal primary wavefunction be delta-function-normalizable with respect to the Klein-Gordon inner product \eqref{KGip}.  The Klein-Gordon inner product of the massless conformal primary wavefunctions with $\Delta  =  c+i\nu$ is
\begin{align}
&\left( \varphi^\pm_{c + i\nu_1} (X^\mu;\vec w_1) \,,\,
\varphi^\pm_{c+i\nu_2}(X^\mu;\vec w_2) \right)\notag\\
&
= \pm 2(2\pi)^{d+1} 
\int_0^\infty d\omega_1\, \omega_1^{c-1+i\nu_1} \,
\int_0^\infty d\omega_2\, \omega_2^{c-1-i\nu_2}\,
\omega_1 (1+|\vec w_1|^2)\,
\delta^{(d+1)}( \omega_1 q^i (\vec w_1) - \omega_2 q^i(\vec w_2))\notag\\
&=\pm 4\pi^{d+1} 
\delta^{(d)}\left( 
\vec w_1-\vec w_2\right)\, 
\int_0^\infty d\omega_2 \, \omega_2^{2c-d +i \nu _1- i\nu_2 -1}
\,.
\end{align}
The $\omega_2$ integral is divergent unless  $c=\frac d2$,  in which case,
\begin{align}\label{m02pt}
\int_0^\infty d\omega \, \omega^{i \nu -1} 
= 2\pi \delta(\nu)\,.
\end{align}
The Klein-Gordon inner product when $c=\frac d2$ is  (with $\nu_1,\nu_2\in\mathbb{R}$)
\begin{align}\label{m0ip}
\left( \varphi^\pm_{\frac d2 + i\nu_1} (X^\mu;\vec w_1) \,,\,
\varphi^\pm_{\frac d2+i\nu_2}(X^\mu;\vec w_2) \right)
=\pm 8\pi^{d+2} 
\,\delta(\nu_1-\nu_2)\,
\delta^{(d)}\left( 
\vec w_1-\vec w_2\right)
\,.
\end{align}
Thus, the massless conformal primary wavefunctions $\varphi^\pm_\Delta$ are delta-function-normalizable  if the conformal dimensions are chosen to be $\Delta= \frac d2 + i\nu$ with $\nu\in\mathbb{R}$, which are again the principal continuous series representations of $SO(1,d+1)$.  The same conclusion was  reached from studying the $AdS$ holography on each hyperbolic slice of  Minkowski space  \cite{deBoer:2003vf,Cheung:2016iub}.

Notice that since the constant wavefunction is not spanned by the outgoing (incoming) conformal primary wavefunctions, the inner products of the latter are strictly positive- (negative-) definite as shown in \eqref{m0ip}.  

The Klein-Gordon inner product further implies that the massless conformal primary wavefunctions with different $\nu\in\mathbb{R}$ are orthogonal to each other.  In particular, $\varphi_{\frac d2+i\nu}$ is \textit{not} linearly related to $\varphi_{\frac d2 -i\nu}$, in contrast to the massive case \eqref{shadowCPW}.  Instead, using \eqref{david}, the shadow of 
$\varphi_{\frac d2+i\nu}$ is 
\begin{align}
\widetilde{ \varphi^\pm_{\frac d2+i\nu}} (X^\mu ;\vec w) 
&= {  \Gamma(\frac d2+i\nu)\over \pi^{\frac d2} \Gamma(i\nu)}
\int d^d \vec z {1\over |\vec z-\vec w|^{2(\frac d2-i\nu)}}
\varphi^\pm_{\frac d2+i\nu}  (X^\mu ;\vec z)  \notag\\
&= (\mp i)^{\frac d2+i\nu} \Gamma(\frac d2 +i\nu)
 \, 
{(-X^2)^{-i\nu} \over (-q(\vec w)\cdot X\mp i\epsilon)^{\frac d2-i\nu}} \,,
\end{align}
which, up to a normalization constant, is $(-X^2)^{-i\nu}\varphi_{\frac d2 -i\nu}^\pm(X^\mu;\vec w)$.  The Klein-Gordon inner product between $\varphi^\pm_{\frac d2+i\nu}$ and its shadow $\widetilde {\varphi^\pm_{\frac d2+i\nu}}$ is then a power law term as  a CFT two-point function:
\begin{align}
\left( \varphi^\pm_{\frac d2 + i\nu_1} (X^\mu;\vec w_1) \,,\,
\widetilde{\varphi^\pm_{\frac d2+i\nu_2}}(X^\mu;\vec w_2) \right)
=\pm 8\pi^{\frac d2+2} 
{\Gamma(\frac d2-i\nu_1)\over \Gamma(-i\nu_1)}
\,\delta(\nu_1-\nu_2)\,
{1\over 
|\vec w_1-\vec w_2|^{2(\frac d2+i\nu_1)}}
\,.
\end{align}

We  conclude that the massless conformal primary bases $\mathcal{B}^\pm_{m=0}$ for the outgoing $(+)$ and  incoming $(-)$, delta-function-normalizable solutions of the massless Klein-Gordon equation are
\begin{align}\label{masslessbasis}
\boxed{\,
\mathcal{B}^\pm _{m=0}=\left\{\,
  \varphi^\pm _{\frac d2 +i\nu}(X^\mu;\vec w) \,\Big| \, \nu\in\mathbb{R}\,,  \vec w\in \mathbb{R}^d \, \right\}  \,
}\,.
\end{align}
  We compare the plane wave bases with the massless conformal primary bases in Table \ref{table:massless}.  We emphasize again that the constant wavefunction is excluded from our definition of either the outgoing or the incoming solution space.  In particular, the constant wavefunction is not spanned by the conformal primary wavefunctions.

Alternatively, the shadows of the bases \eqref{masslessbasis} are equally good conformal primary bases for the outgoing/incoming massless Klein-Gordon solutions:
\begin{align}\label{masslessshadowbasis}
\boxed{
\mathcal{\widetilde B}_{m=0}^\pm=  \left\{\,
\widetilde{ \varphi^\pm _{\frac d2 +i\nu} }(X^\mu;\vec w) \,  \Big|\,  \nu\in \mathbb{R}, \vec w\in \mathbb{R}^d \,\right \}\, }\,.
\end{align}
We have thus identified a pair of bases $\mathcal{B}_{m=0}^+$ and $\mathcal{\widetilde B}_{m=0}^+$ for the outgoing, normalizable solutions of the massless Klein-Gordon equation. Similarly, we have identified a pair of bases $\mathcal{B}_{m=0}^-$ and $\mathcal{\widetilde B}_{m=0}^-$ for the incoming, normalizable solutions of the massless Klein-Gordon equation.  The four bases are related by complex conjugation and shadow transform as below:
\begin{align}
\begin{split}
&\text{Outgoing}:~~~~\mathcal{B}_{m=0}^+   ~~~~ \underset{\text{shadow}}{\longleftrightarrow}~~~~ \mathcal{\widetilde B}_{m=0}^+\\
&~~~~~~~~~~~~~~~~~~~~\Big\updownarrow_{\text{c.c}} ~~~~~~~~~~~~~~~~~~\Big\updownarrow_{\text{c.c}} \\
&\text{Incoming}:~~~~\mathcal{ B}_{m=0}^-   ~~~~ \underset{\text{shadow}}{\longleftrightarrow}~~~~  \mathcal{\widetilde B}_{m=0}^-
\end{split}
\end{align}

\section{Photons}\label{sec:photon}

In this and the following section we present a detailed discussion of  massless conformal primary wavefunctions with spin.  A massless state in $\mathbb{R}^{1,d+1}$ sits in a representation  of the massless little group $SO(d)$, the spin or helicity.  For a spinning conformal primary wavefunction, this spin gets interpreted as that of a conformal primary in $d$ dimensions.  For example, an  outgoing conformal primary wavefunctions in $\mathbb{R}^{1,3}$ with positive helicity transforms as a spin $+1$ conformal primary in two dimensions under $SL(2,\mathbb{C})$.

The spin-one massless conformal primary wavefunction in $(3+1)$ dimensions has been constructed in \cite{Cheung:2016iub}. In Section \ref{sec:spin1cpw} we review this construction and extend it to general spacetime dimensions.  In Section \ref{sec:gauge} we discuss how conformal covariance fixes a particular gauge choice for the conformal primary wavefunctions. We also discuss conformal primary wavefunctions that happen to be  pure gauge.    In Section \ref{sec:spin1basis},  we show that spin-one conformal primary wavefunctions on the principal continuous series $\Delta\in\frac d2+i\mathbb{R}$ are normalizable and span the space of   plane wave solutions to the Maxwell equation.

Since there are no propagating degrees of freedom for a spin-one field in $(1+1)$ spacetime dimensions, we will assume $d\ge1$ in this section.

\subsection{Massless Spin-One Conformal Primary Wavefunctions in General Dimensions}\label{sec:spin1cpw}

The defining properties of the outgoing (+) and incoming ($-$) \textit{massless spin-one conformal primary wavefunction}  $A^{\Delta\pm}_{\mu a}(X^\mu;\vec w)$ in $\mathbb{R}^{1,d+1}$ are ($\mu=0,1,\cdots ,d+1$ and $a=1,\cdots ,d$):
\begin{enumerate}
\item It satisfies the  $(d+2)$-dimensional Maxwell equation,
\begin{align}
\left(  {\partial\over \partial X^\sigma} {\partial\over \partial X_\sigma} \delta_{\nu }^\mu   - {\partial\over \partial X^\nu}{\partial\over \partial X_\mu}\right)    A^{\Delta \pm}_{\mu a} (X^\rho;\vec w) =0\,.
\end{align}
\item It transforms both as a $(d+2)$-dimensional vector  and a $d$-dimensional spin-one conformal primary  with conformal dimension $\Delta$ under an $SO(1,d+1)$ Lorentz transformation:
\begin{align}\label{covarianceV}
&A^{\Delta \pm}_{\mu a } \left(\Lambda^\rho_{~\nu} X^\nu ; \vec w\,'(\vec w)\right)
=
{\partial {w}^b \over \partial {w'}^a}
\left| {\partial \vec w'\over \partial \vec w }\right|^{- (\Delta-1)/d}
\, \Lambda_\mu ^{~\sigma} A^{\Delta\pm}_{\sigma b}(X^\rho;\vec w)\,,
\end{align}
where $\vec w\,'(\vec w)$ is an element of $SO(1,d+1)$ defined in \eqref{CT} and $\Lambda^\mu_{~\nu}$ is the associated group element in the $(d+2)$-dimensional representation.
\end{enumerate}
As usual, a solution to the Maxwell equation is subject to the ambiguity of  gauge transformations.  We will return to this in Section \ref{sec:gauge}.

Similar to the scalar massless conformal primary wavefunction, its spin-one analog has been obtained \cite{Cheung:2016iub} from the  spin-one bulk-to-boundary propagator in $H_{d+1}$ \cite{Witten:1998qj,Freedman:1998tz}. The uplift of the latter in the embedding space $\mathbb{R}^{1,d+1}$ with conformal dimension $\Delta$ is \cite{Costa:2014kfa}:
\begin{align}
G^\Delta _{\mu ; \nu }(\hat p ;q )  = {(- q\cdot \hat p)\eta_{\mu\nu}  +q_\mu \hat p_\nu \over (-q\cdot \hat p)^{\Delta+1}}\,,
\end{align}
where $\hat p^\mu$ is a unit timelike vector and $q^\nu$ is a null vector \eqref{qmap}, both living in $\mathbb{R}^{1,d+1}$.  The uplifted bulk-to-boundary propagator satisfies the following two transversality conditions:
\begin{align}\label{transverse}
\hat p^\mu G^\Delta _{\mu ; \nu }(\hat p ;q ) =0\,,~~~~~~~
q^\nu G^\Delta _{\mu ; \nu }(\hat p ;q ) =0\,.
\end{align}
Under $SO(1,d+1)$, $G^\Delta_{\mu;\nu}$ transforms as a $(d+2)$-dimensional rank-two tensor but a scalar conformal primary in $d$ dimensions:
\begin{align}
G^\Delta_{\mu; \nu}  (\Lambda \hat p ; \Lambda q)
= \left| {\partial \vec w\,'\over \partial \vec w} \right|^{-\Delta/d} \Lambda^\rho_{~\mu} \Lambda^\sigma_{~\nu}\,
G^\Delta_{\rho ;\sigma} (\hat p ; q)\,.
\end{align} 
This is not quite what we want for the conformal primary wavefunction \eqref{covarianceV}, but its projection
\begin{align}\label{projected}
{\p q^\nu \over \p w^a} G^\Delta_{\mu;\nu} (\hat p ; q)
\end{align}
 on the second index $\nu$ does have the desired conformal covariance \eqref{covarianceV}.   This can be shown by the second transversality condition in \eqref{transverse} and the $SO(1,d+1)$ transformation of $\p_a q^\mu  \equiv {\p\over \p w^a } q^\mu(\vec w)$:
 \begin{align}
 \p_{a'} q^\mu(\vec w\,')  ={\p w^b\over \p {w'}^a}  \left| {\p \vec w\,'\over \p \vec w}\right|^{1/d}
 \Lambda^\mu_{~\nu} \p_b q^\nu(\vec w)
 +{\p w^b \over \p{w'}^a} \, \p_b \left( \left| {\p \vec w\,'\over \p \vec w}\right|^{1/d}
\right) \Lambda^\mu_{~\nu}q^\nu(\vec w)\,.
 \end{align}
 We therefore identify the massless spin-one conformal primary wavefunction as the bulk-to-boundary propagator \eqref{projected} with $\hat p$ replaced by $X^\mu$ \cite{Cheung:2016iub}:
 \begin{align}\label{spin1cpw}
& \boxed{\, 
A_{\mu a}^{\Delta, \pm } (X^\mu ; \vec w)  = {\partial_a q_\mu \over (- q\cdot X \mp i \epsilon)^\Delta} +  {\partial_a q\cdot X\over (-q\cdot X \mp i\epsilon)^{\Delta+1} } q_\mu \,}\notag\\
&~~~~~~~~~~~~~~~~~\,=-{1\over (-q\cdot X\mp i\epsilon)^{\Delta-1}}
{\p \over \p X^\mu } {\p \over \p w^a } 
\log (-q\cdot X\mp i\epsilon)\,.
\end{align}
 It is  straightforward to check that  \eqref{spin1cpw} indeed satisfies the Maxwell equation.

An equally good spin-one conformal primary wavefunction  would be the shadow $\widetilde{A_{\mu a}^{\Delta, \pm }}$ of \eqref{spin1cpw}.  The shadow transform \eqref{spinshadow} for the uplifted wavefunction $A^{\Delta }_{\mu \nu} (X^\mu;\vec w) = G^{\Delta}_{\mu; \nu}(X; q)$ is\footnote{For notational simplicity, we drop the $i\epsilon$ terms and the $\pm$ labels for the outgoing/incoming wavefunctions in this calculation.} 
\begin{align}
\widetilde{A ^\Delta _{\mu\nu}}(X;\vec w)
= {\Gamma(\Delta+1)\over \pi^{d\over2} (\Delta-1) \Gamma(\Delta-\frac d2)}
\int d^d \vec w\,'
{
 (-\frac 12q\cdot q')\delta^\rho_{\nu}
 +\frac12 q'_\nu q^\rho
   \over (-\frac 12q\cdot q')^{d-\Delta+1}}
   {(-q'\cdot X)\eta_{\mu\rho} +q'_\mu X_\rho \over(-q'\cdot X)^{\Delta+1}}\,.
\end{align}
Using  \eqref{david}, the above integral (excluding the prefactor ${\Gamma(\Delta+1)\over  \pi^{d\over2} (\Delta-1) \Gamma(\Delta-\frac d2)}$) can be computed to be
\begin{align}\label{shadowintegral}
&
   { \pi^{d\over2} \Gamma(\Delta-\frac d2) \over\Gamma(\Delta+1)}
   { (-X^2)^{\frac d2 -\Delta} \over (-q\cdot X)^{d-\Delta+2} }\\
   &\times
 \left[
(\Delta-1) (-q\cdot X) \left(
 \eta_{\mu\nu} (-q\cdot X) 
+q_\mu X_\nu\right)
- (d+1-\Delta)q_\nu  \left(
q_\mu  X^2
- X_\mu (q\cdot X)
\right)
\right]\,.\notag
\end{align}
The second term in the bracket drops out after the projection  \eqref{projection}. In the end, the shadow wavefunction $\widetilde{A_{\mu a}^{\Delta }} = \p_a q^\nu \widetilde{A_{\mu \nu}^{\Delta }} $ is simply:
\begin{align}\label{spin1shadow}
\widetilde{A_{\mu a}^{\Delta, \pm }} (X^\mu ; \vec w)  = (-X^2)^{\frac d2-\Delta} A_{\mu a}^{d-\Delta, \pm } (X^\mu ; \vec w) \,.
\end{align}
One can easily show that the shadow wavefunction satisfies the two defining properties of massless spin-one conformal primary wavefunctions as well. 
Similar to the massless scalar case, for general $\Delta$, the shadow transform does not take the conformal primary wavefunction $A_{\mu a}^{\Delta, \pm }$ back to itself with the shadow dimension $d-\Delta$, but to a different wavefunction.

\subsection{Gauge Symmetry}\label{sec:gauge}

Let us  discuss the role of gauge symmetry in $(d+2)$ dimensions.  A general gauge transformation has no nice conformal property and thus  spoils the conformal covariance of conformal primary wavefunctions.  In other words,  conformal covariance \eqref{covarianceV} fixes a gauge for the conformal primary wavefunction \eqref{spin1cpw}.  Indeed, \eqref{spin1cpw} satisfies both the radial gauge and the Lorenz gauge conditions:
\begin{align}
X^\mu A_{\mu a}^{\Delta, \pm }(X^\mu ; \vec w)=0\,,~~~~~ \p^\mu A_{\mu a}^{\Delta, \pm }(X^\mu ; \vec w)=0\,.
\end{align}
 The radial gauge condition $X^\mu A_\mu=0$ comes from the first transversality condition for the bulk-to-boundary propagator in \eqref{transverse}.  Note that for any \textit{on-shell} massless spin-one wavefunction, i.e. a solution to the Maxwell equation, it is always possible to perform a gauge transformation such that it satisfies \textit{both} the radial gauge and the Lorenz gauge conditions \cite{Magliaro:2007qr}.\footnote{This is similar to the fact that  temporal gauge $A^0=0$ and Coulomb gauge $\p^i A_i=0$ can be imposed at the same time for a solution to the Maxwell equation.}   Under these gauge conditions, the Maxwell equation simplifies to
\begin{align}
\p_\rho \p^\rho A_{\mu a}^{\Delta, \pm }(X^\mu ; \vec w)=0\,.
\end{align}
All these properties also apply to the shadow wavefunction \eqref{spin1shadow}.

One natural question is whether the conformal primary wavefunction \eqref{spin1cpw} or its shadow is ever a pure gauge in $(d+2)$ dimensions.  A short calculation of their field strengths shows that, for any $d$,  $A^{\Delta , \pm}_{\mu a}(X^\mu;\vec w)$ is a pure gauge only if $\Delta =1$:
\begin{align}\label{puregauge}
A_{\mu a }^{\Delta=1, \pm} = -{\partial \over \partial X^\mu}  {\p \over \p w^a } 
\log (-q\cdot X\mp i\epsilon) \,.
\end{align}
 This pure gauge  has been studied in the context of soft theorems in $(3+1)$ dimensions in \cite{Cheung:2016iub}. Incidentally, $\Delta=1$ is the shadow dimension (i.e. $\Delta \to d-\Delta$) of a conserved spin-one current in a $d$-dimensional CFT. 

What about the shadow conformal primary wavefunctions \eqref{spin1shadow}?  The shadow transform \eqref{spinshadow} commutes with the $X^\mu$ derivative, and hence, at least naively, should map one pure gauge to another.  This expectation, however, suffers from an important subtlety that we will describe below. 
 The shadow transform   \eqref{spinshadow} is strictly speaking not defined for $\Delta=1$,  in which case \eqref{spin1cpw} is a pure gauge. 
  Indeed, from \eqref{shadowintegral} with $\Delta=1$, we see that the shadow integral after the projection \eqref{projection} is either zero or singular for integer $d$, and is accompanied by a singular  prefactor $\Gamma(\Delta+1)/\pi^{d\over 2}(\Delta-1)\Gamma(\Delta-\frac d2)$ in our normalization for the shadow transform.
In practice, we  first obtain the expression \eqref{spin1shadow} by assuming a generic value of $\Delta$, and only then we analytic continue it  to  all $\Delta$.
  
Even though the shadow wavefunctions \eqref{spin1shadow} satisfy all the required properties of conformal primary wavefunctions, the analytic continuation in $\Delta$ mentioned above generally does not preserve the pure gauge condition.   The naive expectation that the shadow wavefunction with conformal dimension $d-1$ is a pure gauge in general spacetime dimensions is falsified by a direct calculation of the field strength of \eqref{spin1shadow}.  Furthermore,  one can show that the shadow wavefunction \eqref{spin1shadow} is never a pure gauge if $d\neq2$.

There is one exception in $(3+1)$ spacetime dimensions where the shadow wavefunction can be a pure gauge.  When $d=2$, the shadow wavefunction \eqref{spin1shadow} happens to be the same as the pure gauge wavefunction \eqref{puregauge}:
\begin{align}\label{spin1selfshadow}
\widetilde{A^{\Delta=1 , \pm}_{\mu a}}(X;\vec w)=A^{\Delta=1 , \pm}_{\mu a}(X;\vec w)~~~~~~(d=2)\,,
\end{align}
and is thus a pure gauge as well.  
We summarize spin-one conformal primary wavefunctions $A_{\mu a}^{\Delta,\pm}$ and their shadow $\widetilde{A_{\mu a}^{\Delta,\pm}}$ that are pure gauge in Table \ref{table:puregauge}.

\begin{table}[h!]
\centering
\begin{tabular}{|c|c|c|c|}
\hline
 & ~~~~$d=2$~ ~~~
&~~~ $d\neq 2$~ ~~\\
\hline
&&\\
~~~$A_{\mu a}^{\Delta,\pm}$~~~&  
&$\Delta=1$ \\
&$~  \Delta=1~$&\\
$\widetilde{A_{\mu a}^{d-\Delta,\pm}}$ & &$ \times$\\
&&\\
\hline
\end{tabular}
\caption{Spin-one conformal primary wavefunctions and their shadows that are pure gauge in $\mathbb{R}^{1,d+1}$.  For $d=2$, the conformal primary  wavefunction  $A^{\Delta=1,\pm} _{\mu a}$ with $\Delta=1$ is identical to its (formal) shadow $\widetilde{ A^{\Delta=1,\pm} _{\mu a}}$ \eqref{spin1selfshadow}, so we place $\Delta=1$ in the middle between the two rows.}\label{table:puregauge}
\end{table}

\subsection{Mellin Transform}\label{sec:spin1basis}

In this section we will determine the range of $\Delta$ such that the conformal primary wavefunctions \eqref{spin1cpw} are delta-function-normalizable with respect to a certain norm and span the  plane wave solutions to the Maxwell equation.

Let us first review   massless spin-one on-shell wavefunctions in momentum space.  We will be working in Lorenz gauge:
\begin{align}
\p^\mu A_\mu(X)=0\,.
\end{align}
In this gauge, the Maxwell equation reduces to $\p^2 A_\mu=0$. The outgoing and incoming plane waves are  $\epsilon_{\mu } (k) e^{\pm i k\cdot X}$  where $k$ is a null momentum and $\epsilon_{\mu }(k)$ is a polarization vector satisfying $k^\mu\epsilon_{\mu }(k)=0$. 
The residual gauge symmetry preserving the  Lorenz gauge condition is
\begin{align}\label{xresidual}
A_\mu(X) \rightarrow A_\mu(X) +\p_\mu \alpha(X)\,,~~~~~\p^2 \alpha(X)=0\,.
\end{align}
In momentum space, this residual gauge symmetry shifts the polarization vector by $\epsilon_{\mu }(k ) \rightarrow \epsilon_{\mu }(k)+ Ck_\mu$, where $C$ is any constant.  
 We can fix this residual gauge symmetry by choosing the polarization vectors to be $ \p q_\mu(\vec w)/\p w^a$ \eqref{daq}:
  \begin{align}\label{spin1pw}
 \p_a q_\mu\, e^{\pm i\omega q\cdot X}\,,
\end{align}
 where  we have parametrized a null vector $k^\mu$ as $k^\mu = \omega q^\mu(\vec w)$ with $q^\mu(\vec w)$ given in \eqref{qmap} and $\omega>0$.  Here $a=1,\cdots,d$.

There is an inner product on the space of complex solutions, modulo gauge transformations that fall off sufficiently fast at infinity, to the Maxwell equation \cite{Ashtekar:1987tt,Crnkovic:1986ex,Lee:1990nz,Wald:1999wa}:
 \begin{align}\label{spin1ip}
 (A_\mu , A'_{\mu'}) =- i \int d^{d+1}X^i \,\left[ A^{\rho} {F
'}_{0\rho}^{*} 
 -{A'}^{ \rho *} F_{0\rho}
 \right] \,.
 \end{align}
 Using the Maxwell equation $\p^\nu F_{\mu\nu}=0$, one can show that the above inner product does not depend on the choice of the Cauchy surface we integrate over. Furthermore, the integrand is gauge invariant up to a total derivative \cite{Zuckerman:1989cx,Barnich:2001jy,Avery:2015rga}.\footnote{Large gauge transformations generally have  nontrivial inner products with other on-shell wavefunctions because of this boundary term, and thus should be regarded as nontrivial elements in the solution space of the Maxwell equation.  In the following we will only construct conformal primary wavefunctions that span the non-zero energy plane waves \eqref{spin1pw}.}  The plane waves \eqref{spin1pw} are  delta-function-normalizable with respect to this inner product:
 \begin{align}
 \left( \p_{a} q_{\mu}\, e^{\pm i\omega q\cdot X} ,  
 \p_{b} q'_{\mu}\, e^{\pm i\omega' q'\cdot X} \right)
 = \pm 8(2\pi)^{d+1} \,\delta_{ab} \, \omega q^0 \, \delta^{(d+1)}(\omega q^i - \omega'{q'}^i) \,.
 \end{align}

Let us now switch gears to conformal primary wavefunctions.  It will prove convenient to choose a particular gauge representative of \eqref{spin1cpw}.  As discussed in Section \ref{sec:gauge},  conformal covariance \eqref{covarianceV}  fixes  the conformal primary wavefunctions \eqref{spin1cpw} to be in radial gauge $X^\mu A_\mu=0$ and  Lorenz gauge $\p^\mu A_\mu=0$ at the same time. However, for the purpose of computing any gauge invariant physical observables such as scattering amplitudes,  we can work with any wavefunction that is  equivalent to \eqref{spin1cpw} by a  gauge transformation, and still obtain a conformally covariant answer at the end of the day.  

A convenient gauge representative of the conformal primary wavefunction is:
\begin{align}\label{spin1cpw2}
\varphi_{\mu a}^{\Delta , \pm } (X^\mu ; \vec w) =(\mp i)^\Delta\Gamma(\Delta)\, {\p_a q_\mu \over (-q\cdot X\mp i\epsilon)^\Delta}\,,
\end{align}
which satisfies the Lorenz gauge condition but not the radial gauge condition.  
Up to a normalization factor, $\varphi_{\mu a}^{\Delta , \pm } $ is gauge equivalent to the conformal primary wavefunction \eqref{spin1cpw} by the following pure gauge:
\begin{align}
{\p \over \p X^\mu} \left( {\p_a q\cdot X \over (-q\cdot X\mp i\epsilon)^{\Delta}}\right)\,.
\end{align}
In fact, since the gauge parameter $\alpha={\p_a q\cdot X \over (-q\cdot X\mp i\epsilon)^{\Delta}}$ satisfies $\p^2\alpha=0$, it is a  residual gauge transformation \eqref{xresidual} preserving the Lorenz gauge condition.  
It follows  that even though \eqref{spin1cpw2} does not transform covariantly under $SO(1,d+1)$ as in \eqref{covarianceV}, the non-covariant terms are pure residual gauge \eqref{xresidual}.  For this reason we will still call $\varphi^{\Delta, \pm}_{\mu a}$ a spin-one conformal primary wavefunction.

The particular gauge representative \eqref{spin1cpw2}  is chosen such that it is related to the plane wave \eqref{spin1pw} by a Mellin transform as in the case of massless scalars,
\begin{align}\label{spin1mellin}
\varphi_{\mu a}^{\Delta , \pm } (X^\mu ; \vec w) = 
\int _0^\infty d\omega \omega^{\Delta-1} \, \left(
\p_a q_\mu \,
e^{\pm i \omega q\cdot X-\epsilon \omega}\right)
\,.
\end{align}
It follows that the  same argument  in Section \ref{sec:m0basis} can be directly borrowed for the spin-one case.  We  conclude that spin-one conformal primary wavefunctions on the principal continuous series $\Delta\in \frac d2 +i\mathbb{R}$ are delta-function-normalizable with respect to \eqref{spin1ip} and span the  plane wave solutions \eqref{spin1pw} of the Maxwell equation.
 Similar to the massless scalar case, another equally interesting space of on-shell wavefunctions with the same property is the shadow of \eqref{spin1cpw2} on the principal continuous series.
Given a gluon scattering amplitude, the transition from momentum space to the space of conformal primary wavefunctions is then implemented by a Mellin transform \eqref{spin1mellin} (or plus  a shadow transform \eqref{spinshadow}) on each external gluon particle.

\section{Gravitons}\label{sec:graviton}

We now turn to massless spin-two conformal primary  wavefunctions.  In  Section \ref{sec:spin2cpw} we construct solutions to the $(d+2)$-dimensional vacuum linearized Einstein equation that transform as spin-two conformal primaries in $d$ dimensions.  In Section \ref{sec:diff} we identify pure diffeomorphisms that are also conformal primary wavefunctions.  In Section \ref{sec:spin2basis} we show that again the spin-two conformal primary wavefunctions on the principal continuous series are normalizable (with respect to \eqref{spin2ip}) and span the plane wave solutions of the linearized Einstein equation.

 Since there are no propagating degrees of freedom in flat space for gravitons below $(3+1)$ dimensions, we will assume $d\ge 2$ in this section.

\subsection{Massless Spin-Two Conformal Primary Wavefunctions in General Dimensions}\label{sec:spin2cpw}

The defining properties for the outgoing (+) and incoming $(-)$ \textit{massless spin-two conformal primary wavefunction} $h^{\Delta, \pm}_{\mu_1 \mu_2 ; a_1 a_2}(X^\mu; \vec w)$ in $\mathbb{R}^{1,d+1}$ are
\begin{itemize}
\item It is symmetric both in the $(d+2)$- and $d$-dimensional vector indices and traceless in the latter:
\begin{align}\label{trless}
\begin{split}
&h^{\Delta, \pm}_{\mu_1 \mu_2 ; a_1 a_2}=h^{\Delta, \pm}_{\mu_2 \mu_1 ; a_1 a_2}\,,~~~~
\\
&h^{\Delta, \pm}_{\mu_1 \mu_2 ; a_1 a_2}=h^{\Delta, \pm}_{\mu_1 \mu_2 ; a_2 a_1}\,,~~~~
\delta^{a_1a_2}h^{\Delta, \pm}_{\mu_1 \mu_2 ; a_1 a_2}=0\,.
\end{split}
\end{align}
\item It is a solution to the vacuum linearized Einstein equation in flat space:\footnote{For notational simplicity, we omit the superscript $\Delta,\pm$ of the wavefunction in this equation.}
\begin{align}
 \p_\sigma \p_\nu h^{\sigma}_{~\mu ;a_1a_2}
+\p_\sigma \p_\mu h^\sigma_{~\nu;a_1a_2}
-\p_\mu \p_\nu h^{\sigma}_{~\sigma;a_1a_2}
-\p^\rho \p_\rho h_{\mu\nu;a_1a_2}
=0\,.
\end{align}
\item It transforms both as a $(d+2)$-dimensional rank-two tensor  and a $d$-dimensional spin-two conformal primary  with conformal dimension $\Delta$ under an $SO(1,d+1)$ Lorentz transformation:
\begin{align}\label{covarianceh}
&h^{\Delta, \pm}_{\mu_1\mu_2; a_1a_2 } \left(\Lambda^\rho_{~\nu} X^\nu ; \vec w\,'(\vec w)\right)
=
{\partial {w}^{b_1} \over \partial {w'}^{a_1}}
{\partial {w}^{b_2} \over \partial {w'}^{a_2}}
\left| {\partial \vec w'\over \partial \vec w }\right|^{- (\Delta-2)/d}
\, \Lambda_{\mu_1} ^{~\sigma_1}\, \Lambda_{\mu_2} ^{~\sigma_2}
 h^{\Delta,\pm}_{\sigma_1\sigma_2 ;  b_1b_2}(X^\rho;\vec w)\,,
\end{align}
where $\vec w\,'(\vec w)$ is an element of $SO(1,d+1)$ defined in \eqref{CT} and $\Lambda^\mu_{~\nu}$ is the associated group element in the $(d+2)$-dimensional representation.
\end{itemize}

With our experiences from the scalar and spin-one wavefunctions, we can immediately write down the massless spin-two conformal primary wavefunctions from  the $H_{d+1}$ spin-two bulk-to-boundary propagator in the embedding formalism:
\begin{align}\label{spin2cpw}
&\boxed{\,h^{\Delta,\pm}_{\mu_1\mu_2 ;a_1a_2}(X; \vec w)= 
P^{b_1b_2}_{a_1a_2}  \,
 {\left[ (-q\cdot X)\partial_{b_1}q_{\mu_1} +(\partial_{b_1}q\cdot X)q_{\mu_1}\right]
\left[(-q\cdot X)\partial_{b_2}q_{\mu_2} +(\partial_{b_2}q\cdot X)q_{\mu_2}\right]
\over (-q\cdot X)^{\Delta+2}}
\,}
\notag\\
&=P^{b_1b_2}_{a_1a_2}\,
{1\over (-q\cdot X\mp i\epsilon)^{\Delta-2}}
\partial_{b_1}\partial_{\mu_1}
\log (-q\cdot X\mp i\epsilon) \,\,
\partial_{b_2}\partial_{\mu_2}
\log (-q\cdot X\mp i\epsilon) \,,
\end{align}
where $P^{b_1b_2}_{a_1a_2}$ projects a rank-two tensor  to its symmetric traceless components:\footnote{Our convention for symmetrization of indices is $T_{(ab)} \equiv \frac 12(T_{ab}+T_{ba})$.}
\begin{align}
P^{b_1b_2}_{a_1a_2}\equiv 
\delta^{b_1}_{~(a_1} \delta^{b_2}_{~a_2)} - \frac 1d \delta_{a_1a_2}\delta^{b_1b_2}\,.
\end{align}
It is then straightforward to check that \eqref{spin2cpw} satisfies the vacuum linearized Einstein equation. An equally interesting spin-two conformal primary wavefunction is the shadow transform \eqref{spinshadow} of \eqref{spin2cpw}. A direct calculation shows that the shadow primary is:
\begin{align}\label{spin2shadow}
\widetilde{h^{\Delta,\pm}_{\mu_1\mu_2 ;a_1a_2}}(X; \vec w)
=(-X^2)^{\frac d2 -\Delta }h^{d-\Delta,\pm}_{\mu_1\mu_2 ;a_1a_2}(X; \vec w)\,.
\end{align}
One can straightforwardly show that \eqref{spin2shadow} satisfies all the required properties of massless spin-two conformal primary wavefunctions. 
Again for general $\Delta$, the shadow transform does not take the conformal primary wavefunction $h_{\mu_1\mu_2; a_1a_2}^{\Delta, \pm }$ to itself with the shadow conformal dimension $d-\Delta$, but to a different wavefunction.

\subsection{Diffeomorphism}\label{sec:diff}

As in the photon case,  conformal covariance \eqref{covarianceh} picks a particular diffeomorphism choice  for the conformal primary wavefunction. 
It is  easy to check  \eqref{spin2cpw}  is traceless and satisfies the Lorenz\footnote{In general relativity, Lorenz gauge is usually defined as $\p^\mu h_{\mu\nu}-\frac 12 \p_\nu h^\rho_{~\rho}=0$.  Since our wavefunction is traceless, the Lorenz gauge condition reduces to $\p^\mu h_{\mu\nu}=0$.} as well as the radial gauge conditions:
\begin{align}\label{fixdiff}
\eta^{\mu_1\mu_2}h^{\Delta,\pm}_{ \mu_1\mu_2; a_1a_2}
=0\,,~~~~~~\partial^{\mu} h^{\Delta,\pm}_{ \mu\mu_2; a_1a_2}
=0\,,~~~~~~X^\mu h^{\Delta , \pm}_{ \mu\mu_2; a_1a_2}=0\,
.
\end{align}
In fact, any solution to the vacuum linearized Einstein equation is diffeomorphic to another solution that satisfies all three conditions in \eqref{fixdiff} \cite{Magliaro:2007qr}.  In this gauge \eqref{fixdiff}, the vacuum linearized Einstein equation becomes
\begin{align}
\p^\rho \p_\rho \,h^{\Delta,\pm}_{ \mu_1\mu_2; a_1a_2}
(X^\mu; \vec w)=0\,.
\end{align}
All these properties also apply to the shadow wavefunction \eqref{spin2shadow}.

While a general diffeomorphism has no nice conformal covariance, the massless spin-two conformal primary wavefunction $h^{\Delta,\pm}_{\mu_1\mu_2;a_1a_2}$ happens to be a pure diffeomorphism if $\Delta=0$ or $\Delta =1$ for any $d$. In these two cases they can be written as 
\begin{align}\label{purediff0}
&h^{\Delta= 0 ,\pm } _{\mu_1\mu_2 ;a_1a_2}(X^\mu ; \vec w) 
=\partial_{\mu_1} \xi^0_{\mu_2 ;a_1a_2}+\partial_{\mu_2} \xi^0_{\mu_1 ;a_1a_2}\,,  \notag\\
&\xi^0_{\mu ;a_1a_2} 
= \frac 12P^{b_1b_2}_{a_1a_2}\,\,
(q\cdot X\pm i\epsilon) \,
\partial_{b_1} \left[ \, q_\mu \partial_{b_2} \log(-q\cdot X \mp i\epsilon)\,\right]\,,
\end{align}
and
\begin{align}\label{purediff1}
&h^{\Delta= 1,\pm} _{\mu_1\mu_2 ;a_1a_2}(X^\mu ; \vec w) 
=\partial_{\mu_1} \xi^1_{\mu_2 ;a_1a_2}+\partial_{\mu_2} \xi^1_{\mu_1 ;a_1a_2}\,, \notag\\
&\xi^1_{\mu ;a_1a_2} 
= -\frac 14 P^{b_1b_2}_{a_1a_2}  \,\,
 \partial_{b_1} \partial_{b_2} \left[ \, q_\mu \log(-q\cdot X\mp i\epsilon)\,\right]\,.
\end{align}
Incidentally, $\Delta=0$ is the shadow dimension (i.e. $\Delta \to d-\Delta$) of the stress-tensor in a $d$-dimensional CFT.

The shadows of these two wavefunctions \eqref{purediff0} and \eqref{purediff1}, however, are not pure diffeomorphisms in general.  This subtlety is parallel to the one we encountered in the spin-one case in Section \ref{sec:gauge}.  In the spin-two case, the shadow transform \eqref{spinshadow} is not defined  for $\Delta=0,1$,  in which cases the conformal primary wavefunctions \eqref{spin2cpw} reduce to pure diffeomorphisms. 
Nonetheless, the shadow wavefunction can be analytically continued to any value of $\Delta$ from the expression \eqref{spin2shadow}.  The subtlety is that this analytic continuation spoils the pure diffeomorphism condition.  Indeed, by a direct computation of the linearized Riemann curvature  tensor, one can check that the shadow wavefunctions \eqref{spin2shadow} with conformal dimensions $d$ and $d-1$, i.e. the shadow dimensions of 0 and 1, are never pure diffeomorphisms if $d\neq2$.

There is again an exception for spin-two wavefunctions in $(3+1)$ spacetime dimensions.  When $d=2$, the $\Delta=1$ shadow  wavefunction \eqref{spin2shadow} is identical to the pure diffeomorphism wavefunction \eqref{purediff1}, 
\begin{align}\label{spin2selfshadow}
\widetilde {h^{\Delta= 1,\pm} _{\mu_1\mu_2 ;a_1a_2}}(X^\mu ; \vec w) 
= h^{\Delta= 1,\pm} _{\mu_1\mu_2 ;a_1a_2}(X^\mu ; \vec w) 
~~~~~~(d=2)\,,
\end{align}
and is thus a pure diffeomorphism as well.  
Additionally, the shadow wavefunction \eqref{spin2shadow} with conformal dimension 2 is also a pure diffeomorphism:
\begin{align}
&\widetilde{ h^{\Delta= 0,\pm} _{\mu_1\mu_2 ;a_1a_2}}(X^\mu ; \vec w) 
=\partial_{\mu_1} \xi^2_{\mu_2 ;a_1a_2}+\partial_{\mu_2} \xi^2_{\mu_1 ;a_1a_2}\,, ~~~~~~~~(d=2)\\
&\xi^2_{\mu ;a_1a_2} 
=- \frac {1}{24} P^{b_1b_2}_{a_1a_2}  
\left[\, 
 \partial_{b_1} \partial_{b_2} \partial^{c}\left( 
 X^\rho f_{\rho \mu;c} \log(-q\cdot X\mp i\epsilon)
 \right)
 -\frac 12 \partial^c\p_c \p_{b_1}\left( X^\rho f_{\rho\mu ;b_2}\log(-q\cdot X\mp i\epsilon)\right)
\, \right]
\,,\notag
\end{align}
where $f_{\rho \mu ;c} \equiv q_\rho \p_c q_\mu  -q_\mu \p_c q_\rho$. 
In \cite{Cheung:2016iub}, this $\Delta=2$ pure diffeomorphism was discussed in the context of soft graviton theorem in $(3+1)$ spacetime dimensions.  We summarize spin-two conformal primary wavefunctions $h^{\Delta,\pm} _{\mu_1\mu_2 ;a_1a_2}$ and their shadows $\widetilde{ h^{\Delta,\pm} _{\mu_1\mu_2 ;a_1a_2}}$ that are pure diffeomorphisms in Table \ref{table:purediff}.

\begin{table}[h!]
\centering
\begin{tabular}{|c|cc|c|c|}
\hline
 & ~~~~~~~~~~~~~~~~$d=2$&
&~~~ $d\ge 2$~ ~~\\
\hline
&&&\\
~~~$h^{\Delta,\pm} _{\mu_1\mu_2 ;a_1a_2}$~~~&  
&~$\Delta=0$~~~&$\Delta=0,1$ \\
&$ \Delta=1$&&\\
$\widetilde{ h^{d-\Delta,\pm} _{\mu_1\mu_2 ;a_1a_2}}$& &\, $\Delta=2$~~~&$ \times$\\
&&&\\
\hline
\end{tabular}
\caption{Spin-two conformal primary wavefunctions and their shadows that are pure diffeomorphisms in $\mathbb{R}^{1,d+1}$. For $d=2$, the  conformal primary  wavefunction  $h^{\Delta=1,\pm} _{\mu_1\mu_2 ;a_1a_2}$ with $\Delta=1$ is identical to its (formal) shadow $\widetilde{ h^{\Delta=1,\pm} _{\mu_1\mu_2 ;a_1a_2}}$ \eqref{spin2selfshadow}, so we place $\Delta=1$ in the middle between the two rows.}\label{table:purediff}
\end{table}

\subsection{Mellin Transform}\label{sec:spin2basis}

Finally, let us determine the range of the conformal dimension $\Delta$ for the spin-two conformal primary wavefunctions so that they are normalizable with respect to a certain norm and span the  plane wave solutions of the linearized Einstein equation.

As before, we first review solutions in momentum space.  We will be working in Lorenz gauge:
\begin{align}
\p^\mu h_{\mu\nu}-\frac 12 \p_\nu h^\rho_{~\rho}=0\,,
\end{align}
in which the vacuum linearized Einstein equation reduces to $\p^\rho \p_\rho h_{\mu\nu}=0$.  The outgoing and incoming plane waves are $\epsilon_{\mu\nu}(k) e^{\pm ik \cdot X}$ where $k^\mu$ is null and
 $\epsilon_{\mu\nu}(k)$ is a symmetric polarization tensor satisfying $k^\mu \epsilon_{\mu\nu}= \frac 12 k_\nu \epsilon^\mu_{~\mu}$.  The residual diffeomorphisms preserving the Lorenz gauge condition are
\begin{align}\label{spin2residual}
h_{\mu\nu}(X) \rightarrow h_{\mu\nu}(X) + \partial_\mu \xi_\nu(X) + \p_\nu \xi_\mu(X)\,,~~~~~~\p^\rho\p_\rho \xi_\mu (X)=0\,.
\end{align}
In momentum space, the residual diffeomorphisms shift the polarization tensor by $\epsilon_{\mu\nu}(k) \rightarrow \epsilon_{\mu\nu}(k) + k_\mu r_\nu + k_\nu r_\mu$ for any vector $r_\mu$.  Using  these residual diffeomorphisms, we can bring the polarization tensor to a rank-two symmetric traceless tensor of the following form:
\begin{align}\label{spin2pw}
g^\pm_{\mu\nu;a_1a_2}(X;\omega, q)=
P^{b_1b_2}_{a_1a_2} \,\p_{b_1} q_\mu \p _{b_2} q_\nu \, e^{\pm i\omega q\cdot X}\,,
\end{align}
with $\p_aq^\mu$ given by \eqref{daq}. Again we have parametrized a null vector $k^\mu$ as $k^\mu = \omega q^\mu(\vec w)$ with $\omega>0$.

The inner product on the space of complex solutions to the vacuum linearized Einstein equation, modulo diffeomorphisms that fall off sufficiently fast at infinity, is \cite{Ashtekar:1987tt,Crnkovic:1986ex,Lee:1990nz,Wald:1999wa,Hawking:2016sgy}
\begin{align}\label{spin2ip}
\left( h_{\mu\nu} , h'_{\mu'\nu'}\right)
&= -{ i} \int d^{d+1}X^i \Big[\,
h^{\mu\nu} \p_0 {h'}^*_{\mu \nu}  - 2h^{\mu\nu} \p_\mu {h'}^*_{0\nu}
+h\p^\mu {h'}^*_{0\mu } - h\p_0 {h'}^*+ h_{ 0\mu}\p^\mu {h'}^*  \notag\\
& -(h\leftrightarrow {h'}^*)\,
\Big]\,,
\end{align}
where $h= h^\rho_{~\rho}$.  Using the linearized  Einstein equation we can show that the inner product does not depend on the choice of  Cauchy surface. Further, it is invariant under diffeomorphism up to a boundary term \cite{Zuckerman:1989cx,Barnich:2001jy,Avery:2015rga,Hawking:2016sgy}.\footnote{Similar to the  spin-one case, large diffeomorphisms generally have nonzero inner products with other on-shell wavefunctions and should therefore be included as  nontrivial elements in the solution space. We will focus on the non-zero energy plane wave solutions \eqref{spin2pw} to the linearized Einstein equation below.} The inner products between two plane waves are
\begin{align}
&\left( g^\pm_{\mu\nu ;a_1a_2} (X;\omega,q) , g^\pm_{\mu'\nu';a_1'a_2'}(X;\omega' ,q')\right) \notag\\
&~~~~~~~~~~~~~~~~~
=\pm 32 (2\pi)^{d+1}\left( \delta_{a_1 (a_1'} \delta_{a_2')a_2} -\frac 1d \delta_{a_1'a_2'}\delta_{a_1a_2}\right)
\, \omega q^0 \, \delta^{(d+1)} (\omega q^i - \omega' {q'}^i)\,.
\end{align}

We now wish to find the space of conformal primary wavefunctions that  spans the plane wave solutions \eqref{spin2pw}.  We start by considering a particular diffeomorphism representative of the conformal primary wavefunction. Let us consider one of the terms in the conformal primary wavefunction \eqref{spin2cpw}:
\begin{align}\label{spin2cpw2}
\varphi ^{\Delta , \pm }_{\mu_1\mu_2;a_1a_2} (X^\mu ; \vec w)
=(\mp i)^\Delta\Gamma(\Delta)\,
P^{b_1b_2}_{a_1a_2}\, 
{\p_{b_1}q_{\mu_1}  \p_{b_2} q_{\mu_2}  \over (-q\cdot X\mp i\epsilon)^\Delta} \,.
\end{align}
It is traceless and satisfies the Lorenz gauge condition $\p^\mu \varphi_{\mu\mu_2;a_1a_2}^{\Delta,\pm}=0$, but  not the radial gauge condition.  It also satisfies the linearized Einstein equation in this gauge, $\p^\rho \p_\rho \varphi^{\Delta,\pm}_{\mu_1\mu_2;a_1a_2}=0$.  One can straightforwardly show that \eqref{spin2cpw2} differs from the spin-two conformal primary wavefunction \eqref{spin2cpw} by a pure residual diffeomorphism \eqref{spin2residual}.  Indeed, even though \eqref{spin2cpw2} does not transform covariantly as in \eqref{covarianceh}, the non-covariant terms of the transformation  are pure residual diffeomorphisms of the Lorenz gauge condition:
\begin{align}
{ \p_bq_{(\mu_1} q_{\mu_2)}\over (-q\cdot X)^\Delta}
= {1\over \Delta-1} \p_{(\mu_1} \left( {\p_b q_{\mu_2)}\over (-q\cdot X)^{\Delta-1}}\right)
\,,~~~~~~{q_{\mu_1} q_{\mu_2} \over (-q\cdot X)^{\Delta}}
= {1\over \Delta-1} \p_{(\mu_1} \left( { q_{\mu_2)}\over (-q\cdot X)^{\Delta-1}}\right)\,.
\end{align}
  Hence for the purpose of computing any diffeomorphism invariant observables, we can use the wavefunction $\varphi^{\Delta, \pm} _{\mu_1\mu_2;a_1a_2}$  and still reproduce a conformally covariant answer in the end.

The advantage of this diffeomorphism representative  \eqref{spin2cpw2} for the  conformal primary wavefunction is that it is simply the Mellin transform of the pane wave \eqref{spin2pw} as in the massless scalar case:
\begin{align}\label{spin2mellin}
\varphi_{\mu_1\mu_2; a_1a_2}^{\Delta , \pm } (X^\mu ; \vec w) = 
\int _0^\infty d\omega \omega^{\Delta-1} \, 
g_{\mu_1\mu_2; a_1a_2}^{\Delta , \pm } (X^\mu ; \omega, q^\mu)\,.
\end{align}
Following the same argument used in Section \ref{sec:m0basis},  we conclude that   
spin-two conformal primary wavefunctions  on the principal continuous series $\Delta\in\frac d2+i\mathbb{R}$ are delta-function-normalizable (with respect to \eqref{spin2ip}) and span the  plane wave solutions \eqref{spin2pw} of the vacuum linearized Einstein equation.  Another equally interesting set of on-shell  wavefunctions with the same property is the shadow of \eqref{spin2cpw2}. 
Given a graviton scattering amplitude, the transition from momentum space to the space of conformal primary wavefunctions is again implemented by a Mellin transform \eqref{spin2mellin} (or plus  a shadow transform \eqref{spinshadow}) on each external graviton particle.

\section*{Acknowledgements}
We are especially grateful to Andy Strominger for various enlightening discussions, collaborations on related projects, and detailed comments on a draft.  
We would also like to thank N. Arkani-Hamed, H.-Y. Chen,   X. Dong, D. Harlow, Y.-t. Huang,  D. Jafferis,  Y.-H. Lin, J. Maldacena, M. Mezei, P. Mitra, H. Ooguri, B. Schwab, D. Simmons-Duffin, D. Stanford, A. Volovich,  A. Zhiboedov, and M. Zlotnikov for interesting conversations.    S.P. is supported by the National
Science Foundation and by the Hertz Foundation through a Harold and Ruth Newman Fellowship.   S.H.S. is supported by the National
Science Foundation grant PHY-1606531.

\bibliography{PrincipalSeries}{}
\bibliographystyle{utphys}

\end{document}